\documentclass[structabstract]{aa}
\usepackage{graphicx}
\usepackage{url}
\usepackage{txfonts}
\usepackage{pstricks}
\usepackage{natbib}
\usepackage{subfigure}
\usepackage{color} 

\bibpunct{(}{)}{;}{a}{}{,}
\begin{document}
\title{COSMOGRAIL: the COSmological MOnitoring of GRAvItational Lenses\thanks{Based on observations made with the NASA/ESA HST Hubble Space Telescope, obtained from the data archive at the Space Science Institute, which is operated by AURA, the Association of Universities for Research in Astronomy, Inc., under NASA contract NAS-5-26555.}}
\subtitle{VIII. Deconvolution of high resolution near-IR images and simple mass models for 7 gravitationally lensed quasars}

\titlerunning{COSMOGRAIL: Accurate astrometry and models for 7 lensed quasars}

\author{V. Chantry 
\inst{1}
\thanks{Research Fellow, Belgian National Fund for Scientific Research (FNRS)}
\and D. Sluse
\inst{2}
\thanks{Alexander von Humboldt Fellow}
\and P. Magain 
\inst{1}
}

\institute{Institut d'Astrophysique et de G\' eophysique, Universit\' e de Li\`ege, All\'ee du 6 Ao\^ut, 17, 4000 Sart Tilman (Bat. B5C), Li\`ege 1, Belgium
\email{Virginie.Chantry@ulg.ac.be}
\and
Astronomisches Rechen-Institut am Zentrum f\"ur Astronomie der Universitat at Heidelberg,
M\"onchhofstrasse 12-14, 69120 Heidelberg, Germany
}
\date{}
 
  \abstract
  {}  
{We aim at obtaining very accurate positional constraints for seven gravitationally lensed quasars currently monitored by the COSMOGRAIL collaboration, and shape parameters for the light distribution of the lensing galaxy. We also want to find simple mass models that reproduce the observed configuration and predict time delays. Finally we want to test, for the quads, whether there are clues of astrometric perturbations coming from substructures in the lensing galaxy, preventing from finding a good fit of the simple models.}
{We apply the iterative MCS deconvolution method to near-IR HST archives data of seven gravitationally lensed quasars. This deconvolution method allows us to separate the contribution from the point sources to the one from extended structures such as Einstein rings. That method leads to an accuracy of 1-2 mas on the relative positions of the sources and lens. The limiting factor of the method is the uncertainty on the instrumental geometric distortions. We then compute mass models of the lensing galaxy using state-of-the-art modeling techniques.}
{We obtain relative positions for the lensed images and lens shape parameters of seven lensed quasars: HE~0047-1756, RX~J1131-1231, SDSS~J1138+0314, SDSS~J1155+6346, SDSS~J1226-0006, WFI~J2026-4536 and HS~2209+1914. The lensed image positions are derived with 1-2 mas accuracy. Isothermal and de Vaucouleurs mass models are calculated for the whole sample. The effect of the lens environment on the lens mass models is taken into account with a shear term. Doubly imaged quasars are equally well fitted by each of these models. A large amount of shear is necessary to reproduce SDSS~J1155+6346 and SDSS~J1226-006. In the latter case, we identify a nearby galaxy as the dominant source of shear. The quadruply imaged quasar SDSS~J1138+0314 is well reproduced by simple lens models, which is not the case for the two other quads, RX~J1131-1231 and WFI~J2026-4536. This might be the signature of astrometric perturbations due to massive substructures in the galaxy unaccounted for by the models. Other possible explanations are also presented.}
  {}  
\keywords{Gravitational lensing: astrometry --
          Quasars: general --
	Cosmology: observations --
	  Techniques: image processing   }

\maketitle
%

\section{Introduction}

\citet{Refsdal1964} was the first to state that gravitationally lensed quasars can be very useful for determining parameters of our Universe: combined with a model of the mass distribution in the lensing galaxy, the time delay between different lensed images can lead to the determination of the \textit{Hubble constant}, $H_{0}$. This motivated many of the early lensed quasars studies and time delay measurements campaigns. Unfortunately it quickly became clear that the final estimate of $H_{0}$ is very sensitive to systematic errors on the lens modeling. One way to reduce these systematic errors is to derive accurate relative astrometry of gravitationally lensed images and lens galaxy light profiles based on high resolution frames. A good example of the consequences on $H_{0}$ of accurate astrometry is the quadruply lensed quasar PG~1115+080: \citet{Courbin1997} found stronger constraints on the lensed images and lensing galaxy positions which reduced the degeneracies and the range of acceptable models. This allowed to measure $H_{0}$ with an accuracy two times better than in previous studies. \citet{Keeton1997} also showed that reducing the error on the lens galaxy position of PG~1115+080 from 50 mas to 10 mas could improve the constraints on the lens model and thus on $H_{0}$. Moreover \citet{Lehar2000} highlighted that a poor knowledge in the position of the lens galaxy in two different systems, i.e. B0218+357 and PKS~1830-211, prevented from accurately measuring the Hubble constant. Systematic studies of the effect of the astrometric accuracies on $H_{0}$ are difficicult as they depend on the lens system configuration.

Strong lensing is also a promising tool to estimate the amount (possibly as a function of redshift) of dark matter clumps (hereafter, following other authors, we will call them \textquotedblleft substructures\textquotedblright)  in distant galaxies and compare it to predictions of numerical simulations \citep[see e.g.][]{Zackrisson2009, Koopmans2009b}. The first evidence that strongly lensed quasars are sensitive to substructures in galaxies comes from the so-called \textquotedblleft anomalous flux ratios\textquotedblright: for many systems, the flux ratios between the lensed images deviate from those predicted by simple lens models \citep{Kochanek1991, Mao1998, Dalal2002, Keeton2003}. It has long been thought that substructures only act on the image flux ratios because of the dependence of the latters on the second derivative of the gravitational potential. However, recent works explore two new routes to detect substructures in lensed quasars. One method suggests to use time delay measurements, as shown by \citet{Keeton2009} who have highlighted small changes in time delays because of substructures. Even if these delays are likely to be modified only by a few tenths of a percent, future large monitoring campaigns should allow the detection of the signature of substructures \citep{Moustakas2009}. Another method proposes to detect substructures in the lensing galaxy through their effects on the position of lensed images. The amplitude and probability we should expect for this phenomenon is still debated. On one hand, observable astrometric perturbations should be due to the most massive substructures. But because of the scarcity of high mass dark matter clumps, \citet{Metcalf2001} derived a low probability and, on average, low astrometric perturbations. On the other hand, \citet{Chen2007} showed that lower mass substructures also play a role. Including a large range of sub-halo masses, they find that substructures could induce astrometric perturbations as large as 10 mas \citep[see][ for a more complete review]{Zackrisson2009}. Observationaly, astrometric perturbations caused by substructures were detected in a few systems, the most remarkable ones being MG2016+112 \citep{Koopmans2002, More2009} and B0128+437 \citep{Biggs2004}. In both cases, the anomalies have been unveiled thanks to high resolution radio images. 

Although we cannot yet reach the spatial resolution of the \emph{Very Long Baseline Array} (VLBA) in the optical range, it has been shown by some of us \citep{chantry2007} that a sophisticated deconvolution technique (ISMCS\footnote{Iterative Strategy combined with the MCS deconvolution alogrithm}) applied to \emph{Hubble Space Telescope} (HST) images could lead to relative positions of lensed quasar images with milliarcsecond (mas) accuracy, reducing the error bars by a factor $>$ 2 compared to other techniques. In the present paper, we apply this technique to a sample of seven gravitational lenses without measured time delays. All these systems are photometrically monitored by the COSMOGRAIL{\footnote{COSmological MOnitoring of GRAvItational Lenses; \url{http://www.cosmograil.org}}} collaboration and should get a time delay measurement in the near future. The goals of this paper are twofold. First we want to provide shape parameters for the lensing galaxy and accurate relative astrometry for these systems together with simple lens models. From the latter, prospective time delays are also calculated, complementing time delays predicted with non parametric modeling \citep{Cosmograil4}. Second, we systematically investigate, for quadruply imaged quasars, the ability of simple smooth models to reproduce the image configuration within a few milliarcsec. From this systematic and uniform approach, we want to test whether the actual data show evidence for astrometric perturbations due to substructures. 

The lens sample studied in this paper is composed of 7 different systems without time delay measurements and currently monitored by the COSMOGRAIL collaboration: 4 doubly imaged quasars for which no detailed modeling and/or relative astrometry has ever been published and 3 quadruply imaged quasars, amongst which 2 have not yet been studied in detail either (no modeling, no time delay and/or no lens redshift). The ISMCS deconvolution of the gravitational lenses for which time delays have already been measured will be presented in another paper (Chantry et al., in preparation). 

The studied sample is detailed in Sect. \ref{material7} and the data in Sect. \ref{obs7}, while the image processing technique is explained in Sect. \ref{dec7} along with the results. The modeling strategy is explained in Sect. \ref{model7}. A discussion of the models is presented in Sect. \ref{Discussion7}. We then conclude in Sect. \ref{conc7}.

\section{An overview of our sample}
\label{material7}

Here are the seven gravitationally lensed quasars of our sample, the right ascension and declination being expressed in the J2000 coordinates system:
\begin{itemize}

 \item \textbf{HE~0047-1756 (a)}

This object (RA = $00^{\rm h}50^{\rm m}27^{\rm s}.82$ and DEC = $-17^{\circ}40'08\farcs79$) was discovered by \citet{Wisotzki2000} in the framework of the Hamburg/ESO Survey (HES) for bright quasars, covering the Southern sky. It was later identified by \citet{Wisotzki2004} as a doubly imaged quasar at a redshift of $\rm z_{s}=1.68$. The lens is an elliptical galaxy with a spectroscopic redshift of $\rm z_{l}=0.407\pm0.001$ \citep{Cosmograil3, Ofek2006}. 
\\
 \item \textbf{RX~J1131-1231 (b)}

This quadruply imaged quasar (RA = $11^{\rm h}31^{\rm m}55^{\rm s}.39$ and DEC = $-12^{\circ}31'54\farcs99$) was discovered in \citeyear{Sluse2003} by \citeauthor{Sluse2003} They found a redshift of $\rm z_{l}=0.295\pm0.002$ for the lens while the source lies at $\rm z_{s}=0.657\pm0.001$. Preliminary time delays have been proposed by \citet{Morgan2006} and revised estimates will be published in \citet{Kozlovski}. The system was characterized in details in terms of astrometry and photometry by \citet{Sluse2006}. \citet{Claeskens2006} modeled it and also reconstructed the source which appears to be a Type 1 Seyfert spiral galaxy. A similar work was perfomed by \citet{Brewer2008} using a Bayesian approach. Evidence for substructures in the main lens was searched by \citet{Sugai2007} while \citet{Sluse2007} and \citet{Dai2010} used microlensing to study the quasar source.
\\
 \item \textbf{SDSS~J1138+0314 (c)}

This quadruply imaged object (RA = $11^{\rm h}38^{\rm m}03^{\rm s}.70$ and DEC = $+03^{\circ}14'57\farcs99$) was discovered in \citeyear{Inada2008} in the Sloan Digital Sky Survey (SDSS) by \citeauthor{Inada2008} The redshifts of the quasar and the lens were measured by \citet{Cosmograil3} and are respectively equal to $\rm z_{s}=2.438$ and $\rm z_{l}=0.445\pm0.001$. No detailed modeling and no time delay have ever been published for this system.
\\
 \item \textbf{SDSS~J1155+6346 (d)}

This doubly imaged quasar (RA = $11^{\rm h}55^{\rm m}17^{\rm s}.34$ and DEC = $+63^{\circ}46'22\farcs00$) was discovered by \citet{Pindor2004} in the SDSS data set. They measured the redshifts of the quasar and the lens: $\rm z_{s}=2.888$ and $\rm z_{l}=0.176$. They also found that one of the two images of the quasar is very close to the lensing galaxy (at around 10\% in effective radius off the center of the lens) and is the brightest. That configuration cannot be reproduced by a simple model of mass distribution.
\\
 \item \textbf{SDSS~J1226-0006 (e)}

This system (RA = $12^{\rm h}26^{\rm m}08^{\rm s}.02$ and DEC = $-00^{\circ}06'02\farcs19$) is a doubly imaged quasar discovered in the framework of the SDSS by \citet{Inada2008}. The quasar is located at a redshift of $\rm z_{s}=1.125$. According to \citet{Cosmograil3}, the lens is likely to be an early-type galaxy, with a spectroscopic redshift of $\rm z_{l}=0.516\pm0.001$. This system has no measured time delay, no published relative astrometry and no detailed modeling study.
\\
 \item \textbf{WFI~J2026-4536 (f)}

\citet{Morgan2004} discovered this quadruply imaged quasar (RA = $20^{\rm h}26^{\rm m}10^{\rm s}.43$ and DEC = $-45^{\circ}36'27\farcs10$) during an optical survey using the WFI camera mounted on the MPG/ESO 2.2m telescope operated by the European Southern Observatory (ESO). The redshift of the source is $\rm z_{s}=2.23$. The one of the lens is unknown, although it is clearly visually detected on high resolution images. No time delay has ever been measured but according to \citet{Morgan2004}, the longest one might be of the order of at most a week or two. According to them, the lensed images are likely affected by microlensing.
\\
 \item \textbf{HS~2209+1914 (g)}

This system (RA = $22^{\rm h}11^{\rm m}30^{\rm s}.30$ and DEC = $+19^{\circ}29'12\farcs00$) is a doubly imaged quasar, with $\rm z_{s}=1.07$, discovered during the Hamburg-Cfa Bright Quasar Survey (HS) by \citet{Hagen1999}. They clearly detected the lensing galaxy. Nothing else is available for this system: no time delay, no lens redshift and no modeling.
\\
\end{itemize}

\section{Observational material}
\label{obs7}

The images we analyse were acquired with the camera 2 of NICMOS, i.e. the \textit{Near-Infrared Camera and Multi-Object Spectrometer} (hereafter NIC2)  mounted on the HST. They were all obtained in the framework of the CASTLES project (Cfa-Arizona Space Telescope LEns Survey\footnote{\url{http://www.cfa.harvard.edu/castles/}}, PI: C.S. Kochanek), and are available in the HST archives. The filter used is the F160W which is very close to the $H$-Band filter. It was selected for several reasons: on one hand, the PSF is  well-sampled and only slightly variable across the field; on the other hand, the lensed quasars flux ratios are less affected by microlensing effects, given the size of the quasar at these wavelengths, and dust extinction by the lensing galaxy. Details about the image acquisition are summarized in the first columns of Table \ref{sum_im7}: the name of the object, the date of observation, the number of frames and the total exposure time. All the frames were obtained after the installation of the \textit{NICMOS Cooling System}, or NCS, in 2002. Every image was acquired with dithering and in the MULTIACCUM mode, each one of them being a combination of about twenty subframes. As these objects were all observed between october and december 2003, the pixel size of the detector on the sky does not change from one target to the other, also because the plate scale of NICMOS has become very stable since the installation of the NCS. The values we use were measured during part b of the third Servicing Mission Observatory Verification, SMOV3b, and are the following: x=0.075948\arcsec and y=0.075355\arcsec \citep{Nicmos2007}.

\begin{table*}[ht!]
\centering 
\begin{tabular}{c||ccc|ccc}
\hline
Object & Date of obs. (y-m-d) & \# of frames & Total exp. time & \# of iterations & $\chi^{2}_{r}$ first it. & $\chi^{2}_{r}$ last it.\\  
\hline
\hline
(a) HE~0047-1756 & 2003-12-10 & 4 & 44' & 3 & 60.49 & 1.39\\ 
(b) RX~J1131-1231 & 2003-11-17 & 8 & 89' & 4 & 60.18 & 2.26 \\
(c) SDSS~J1138+0314 & 2003-11-06 & 4 & 44' & 3 & 19.61 & 1.76\\
(d) SDSS~J1155+6346 & 2003-12-12 & 5 & 84' & 4 & 30.15 & 3.33\\
(e) SDSS~J1226-0006 & 2003-11-21 & 4 & 44' & 3 & 20.03 & 1.08 \\
(f) WFI~J2026-4536 & 2003-10-21 & 4 & 46' & 4 & 36.62 & 4.35\\
(g) HS~2209+1914 & 2003-10-14 & 4 & 44' & 4 & 40.39 & 4.2\\
\hline
\end{tabular}
\vspace{0.2cm}
\caption{General information about the acquisition of HST/NIC2 images and the application of ISMCS.}
\label{sum_im7}
\end{table*}

\section{ISMCS on HST/NIC2 images}
\label{dec7}

\subsection{Technique}

To extract accurate spatial and shape parameters from our data, we need a method capable of separating the contributions of the lensed point sources from the ones of the more diffuse components (galaxies, halos, arcs, rings, ...). This is exactly what the MCS deconvolution algorithm \citep{MCS98} provides. One of the advantages of this deconvolution method with respect to other techniques is that it does not violate the sampling theorem. In practice that means that we do not try to fully deconvolve an image in order to obtain an infinite resolution. Instead, we choose a resolution for the final deconvolved image, in our case a Gaussian with 2 pixels \textit{Full-Width-at-Half-Maximum} (FWHM), and we deconvolve our images with a partial \textit{Point Spread Function} or PSF (which gives the total PSF when reconvolved with our 2 pixels FWHM Gaussian). To achieve this task, we need to know very well the shape of the PSF. As the NIC2 field is only 19\farcs2 $\times$ 19\farcs2, we do not have the possibility to use field stars to determine the PSF. Moreover, since the lensed quasar images are contaminated by the lensing galaxy or partial Einstein rings underneath them, we cannot use these images directly to improve our PSF. Instead, we use ISMCS \citep[see][ for further details]{chantry2007}, a special iterative strategy coupled with the MCS algorithm. The HST PSF being quite complex (it includes spike-like features and an intense first Airy ring), we start the deconvolution process using a PSF created by the Tiny Tim software \citep{Tinytim} as a first guess of the true PSF. We improve the Tiny Tim function in adjusting it at the same time on all the point sources of a frame using a technique described in \citet{PSFsimult} which allows to add a numerical component to the input PSF so that it is better adapted to the actual frame. We then obtain a set of modified PSFs which we use to simultaneously deconvolve all the frames at our disposal, with a sampling step two times smaller than the original one. In doing so, we obtain a first approximation of the diffuse background and after reconvolving it to the inital resolution, we subtract it from the original frames and we obtain new ones, partially cleaned from the extended structures. On these modified frames which contain point sources less contaminated by smooth structures, we improve once again our PSFs. This iterative process has to be repeated until the reduced chi square, $\chi^{2}_{r}$ (Eq. \ref{residu_reduit7}), reaches a value close to unity in an area determined by the maximum extension of the residual structures after the first deconvolution, and until the residuals are sufficiently flat (no sharp structure). In practice, we stop when an additional iteration no longer improves the $\chi^{2}_{r}$ (typically, $\Delta\chi^{2}_{r} < 0.2$). For an image with $N$ pixels, the latter is defined as follows:
\begin{equation}
\chi^{2}_{r} = \frac{1}{N} \sum_{\vec{x}} \Bigg( \frac{\mathcal{M}(\vec{x})-\mathcal{D}(\vec{x})}{\sigma(\vec{x})} \Bigg)^{2}
\label{residu_reduit7}
\end{equation}
where $\mathcal{M}(\vec{x})$ is the model reconvolved by the partial PSF, $\mathcal{D}(\vec{x})$ is the observed signal and $\sigma(\vec{x})$ the standard deviation associated to that signal. In practice and as an improvement compared to the process applied to the Cloverleaf in the original paper of \citet{chantry2007}, we noticed that convergence is reached faster when performing a first simultaneous deconvolution of all the frames with Tiny Tim PSFs instead of first trying to improve these PSFs on the unmodified images, i.e. containing the untouched background structures. With this first deconvolution, we obtain a map of the diffuse structures. The latter has then to be cleaned from some artificial ring structures present only to compensate for inaccurate PSFs. It can then be subtracted from the original frames which can then be used to improve a first time the PSFs. That first deconvolution is accounted for as one iteration.

\subsection{Results}

The original frame (combination of all observations), the deconvolved image\footnote{The labels of the lensed images are the same as in previous studies if any.} obtained from the last iteration of ISMCS and the mean residual maps\footnote{The residual map is the image of the difference between the model and the original frame in units of sigma.} after the first and last iterations are displayed for each system in Fig. \ref{dec_NICMOS7}. Both residual maps are expressed in units of $\sigma$ and their color scale ranges from -5 in black to +5 in white. The black rectangle delimits the zone taken into account to estimate the reduced $\chi^{2}$, the orientation of these rectangles being the one of the original frames. We emphasize that the PSF used for the first iteration is the one created by the Tiny Tim software. When we examine the residual maps, the improvement brought by ISMCS is undeniable. Moreover, in most cases (5 amongst 7), the remnant structures underneath the point sources  (on the residual map from the last iteration), are in disagreement with each other, which is the sign of a variable PSF throughout the detector, even on small spatial scales. The number of iterations necessary to reach convergence is shown in the last columns of Table \ref{sum_im7} along with the values of the $\chi^{2}_{r}$ after the first and the last iterations.

The astrometry, corrected from the X/Y scale difference and the distortions of NIC2, and the photometry (Vega system) are shown in Table \ref{astrom7}. The $\pm1\sigma$ error bars were calculated in deconvolving each frame individually at the last iteration and in determining the dispersion around the mean. They are very small because they are inherent to the deconvolution technique: no external systematic error is included in these error bars. To estimate the total error, we compare the spatial extension of each object on the detector to the one of the Cloverleaf (H1413+117). The latter was used as a test of ISMCS in \citet{chantry2007}: in comparing the astrometry of the point sources obtained in two different filters and with two different orientations on the sky, they could estimate the total error to 1 mas, accounting e.g. for a possible remnant distortion in the images. The estimated total errors based on the Cloverleaf are displayed in the fifth column of Table \ref{astrom7}. Of course, as they are based on the maximum extension of the object no matter the direction, they should be considered as upper limits. That is why they are called \textquotedblleft MTE\textquotedblright \ which stands for \textquotedblleft Maximum Total Error\textquotedblright.

Since the total error derived in \citet{chantry2007} for H1413+117 was based on a comparison of the relative positions of the lensed images obtained at different NIR wavelengths and image orientations with the same instrument, we have attempted to get independent estimates based on the comparison of the relative astrometry derived with HST and with high resolution radio data. In a future paper treating the lenses with already measured time delays (Chantry et al., in preparation), we will present HST astrometry for the radio quad JVAS~B1422+231 \citep{Patnaik1992}. To estimate the error affecting our results, we choose one lensed image as astrometric reference and we calculate the distance between it and every other lensed image. We then measure the difference between the distances obtained with our positions and the ones calculated with the radio astrometry of \citet{Patnaik1999}. The scatter of these differences of distance around the mean is about 2.6 mas. Assuming the uncertainty is the same in any direction, we derive an error on the relative astrometry of 1.8 mas in RA and in DEC. This value is larger then the MTE of 1.05 mas derived for B1422+231 from our standard method. This is expected as the radio emission in B1422+231 is slightly extended and is likely not to originate from the accretion disk (observed in the optical range) but rather from a nearby region at the basis of the radio jet. Such an effect, known as core-shift, is observed between two different radio-bands \citep{Porcas2009,Kovalev2008} and may induce astrometric perturbations as large as a few mas on the relative astrometry of lensed quasar images \citep{Mittal2006}. Thus, it appears that the use of radio data as an independant calibrator of the systematic errors at mas accuracy is difficult. It requires the comparison of the relative astrometry, for several objects, between radio and optical wavelengths, a task that is beyond the scope of the present paper.

\begin{table*}
\centering 
\begin{tabular}{c|c||ccc|cc}
\hline
Object & Label & $\Delta RA$ (\arcsec) & $\Delta DEC$ (\arcsec) & MTE (mas) & Magnitude & Flux ratio \\ 
\hline
\hline
(a) HE~0047-1756 &A &  0.        & 0.        & 1.17 & 15.19 $\pm$ 0.01 &  1.\\
&B & 0.2328 $\pm$ 0.0008 & -1.4094 $\pm$ 0.0002 & 1.17 & 16.69 $\pm$ 0.01 & 0.253 $\pm$ 0.002 \\
&G & 0.2390 $\pm$ 0.0022 & -0.8098 $\pm$ 0.0056 & 1.17 & 17.17 $\pm$ 0.02 & /\\
\hline
(b) RX~J1131-1231 & A &  0.        & 0.        & 2.64 & 15.36 $\pm$ 0.01 &  1.\\
& B & 0.0347 $\pm$ 0.0005 & 1.1870 $\pm$ 0.0005 & 2.64 & 15.58 $\pm$ 0.01 & 0.816 $\pm$ 0.003 \\
& C & -0.5920 $\pm$ 0.0007 & -1.1146 $\pm$ 0.0004 & 2.64 & 16.42 $\pm$ 0.01 & 0.374 $\pm$ 0.003 \\
& D & -3.1154 $\pm$ 0.0012 & 0.8801 $\pm$ 0.0013 & 2.64 & 17.76 $\pm$ 0.01 & 0.110 $\pm$ 0.001 \\
& G & -2.0269 $\pm$ 0.0016 & 0.6095 $\pm$ 0.0015 & 2.64 & 15.55 $\pm$ 0.03 & / \\
\hline
(c) SDSS~J1138+0314 & A &  0.        & 0.       & 1.17 & 17.89 $\pm$ 0.01 & 1. \\
& B & -0.1003 $\pm$ 0.0006 & 0.9777 $\pm$ 0.0007 & 1.17 & 19.07 $\pm$ 0.01 & 0.336 $\pm$ 0.004 \\
& C & -1.1791 $\pm$ 0.0003 & 0.8119 $\pm$ 0.0007 & 1.17 & 18.89 $\pm$ 0.01 & 0.400 $\pm$ 0.002 \\
& D & -0.6959 $\pm$ 0.0003 & -0.0551 $\pm$ 0.0003 & 1.17 & 19.02 $\pm$ 0.01 & 0.354 $\pm$ 0.003 \\
& G & -0.4633 $\pm$ 0.0071  &  0.5340  $\pm$ 0.0036 & 1.17 & 17.77 $\pm$ 0.01 &  /\\
\hline
(d) SDSS~J1155+6346 & A &  0.        & 0.       & 1.59 & 16.83 $\pm$ 0.02 &  1.\\
& B & 1.8983 $\pm$ 0.0005 & 0.4052 $\pm$ 0.0005 & 1.59 & 17.87 $\pm$ 0.01 & 0.710 $\pm$ 0.017 \\
& G & 1.6982 $\pm$ 0.0024 & 0.3438 $\pm$ 0.0009 & 1.59 & 15.71 $\pm$ 0.01 &  /\\
\hline
(e) SDSS~J1226-0006 & A &  0.        & 0.        & 1.03 & 17.05 $\pm$ 0.01 &  1.\\
& B & 1.2563 $\pm$ 0.0002 & -0.0550 $\pm$ 0.0007 & 1.03 & 17.80 $\pm$ 0.01 & 0.499 $\pm$ 0.006 \\
& G & 0.4386 $\pm$ 0.0029 &  0.0209 $\pm$ 0.0034 & 1.03 & 17.71 $\pm$ 0.03 &  /\\
\hline
(f) WFI~J2026-4536 & B &  0.        & 0.        & 1.17 & 17.08 $\pm$ 0.01 &  1.\\
& A1 & 0.1613 $\pm$ 0.0007 & -1.4290 $\pm$ 0.0005 & 1.17 & 15.58 $\pm$ 0.01 & 3.988 $\pm$ 0.018 \\
& A2 & 0.4140 $\pm$ 0.0007 & -1.2146 $\pm$ 0.0006 & 1.17 & 16.03 $\pm$ 0.01 & 2.634 $\pm$ 0.017 \\
& C & -0.5721 $\pm$ 0.0006 & -1.0437 $\pm$ 0.0003 & 1.17 & 17.26 $\pm$ 0.01 & 0.851 $\pm$ 0.07 \\
& G & -0.0479 $\pm$ 0.0015 & -0.7916 $\pm$ 0.0015 & 1.17 & 18.94 $\pm$ 0.04 & / \\
\hline
(g) HS~2209+1914 & A &  0.        & 0.        & 0.85 & 14.37 $\pm$ 0.02 & 1. \\
& B & 0.3307 $\pm$ 0.0004 & -0.9863 $\pm$ 0.0010 & 0.85 & 14.63 $\pm$ 0.01 & 0.790 $\pm$ 0.027 \\
& G & 0.2155 $\pm$ 0.0037 & -0.3947 $\pm$ 0.0054 & 0.85 & 21.58 $\pm$ 0.2 & /  \\
\hline
\end{tabular}
\vspace{0.2cm}
\caption{Relative position, maximum total error (\textquotedblleft MTE\textquotedblright), magnitude and flux ratio of the lensed images and lensing galaxy (see Fig. \ref{dec_NICMOS7} for the labels).}
\label{astrom7}
\end{table*}

Together with the point-source deconvolution, we use an analytical model to characterize the lensing galaxy light distribution. To ensure in this case that the maximum amount of light of the galaxy is included in the profile, the deconvolution is performed with no numerical component. Since most of the lensing galaxies are ellipticals, we use a de Vaucouleurs light profile \citep{1948}. This procedure allows us to extract the galaxy shape parameters summarized in Table \ref{gal7}: the position angle or \textquotedblleft PA\textquotedblright \ (orientation in degrees East of North) of the galaxy, its ellipticity ($e=1-b/a$), the effective semi-major and semi-minor axis (resp. $a_{eff}$ and $b_{eff}$). The effective radius $R_{eff}$ is further calculated as being the geometrical mean between the two effective semi-axis \citep{Kochanek2002}. These three quantities are expressed in arcseconds. The $\pm1\sigma$ error bars were also calculated in deconvolving each frame individually and in determining the dispersion around the mean. Let us note that the luminosity of the galaxies displayed in Table \ref{astrom7} is measured in an aperture equal to $R_{eff}$. 

\begin{table*}
\centering 
\begin{tabular}{c||ccccc}
\hline
Object & PA ($^{\circ}$) & $e$ & $a_{eff}$ (\arcsec) & $b_{eff}$ (\arcsec) & $R_{eff}$ (\arcsec) \\ 
\hline
\hline
(a) HE~0047-1756 & 113.8 $\pm$ 5.5 & 0.22 $\pm$ 0.02 & 1.02 $\pm$ 0.03 & 0.81 $\pm$ 0.02 & 0.91 $\pm$ 0.02 \\
(b) RX~J1131-1231 & 108.6 $\pm$ 2.4 & 0.25 $\pm$ 0.04 & 1.25 $\pm$ 0.06 & 0.97 $\pm$ 0.01 & 1.11 $\pm$ 0.03  \\
(c) SDSS~J1138+0314 & 122.7 $\pm$ 6.5 & 0.16 $\pm$ 0.02 & 0.93 $\pm$ 0.04 & 0.79 $\pm$ 0.03 & 0.86 $\pm$ 0.03 \\
(d) SDSS~J1155+6346 & 0.7 $\pm$ 3.4 & 0.15 $\pm$ 0.02 & 1.23 $\pm$ 0.02 & 1.06 $\pm$ 0.01 & 1.14 $\pm$ 0.01  \\
(e) SDSS~J1226-0006 & 45.2 $\pm$ 6.1 & 0.07 $\pm$ 0.02 & 0.72 $\pm$ 0.04 & 0.67 $\pm$ 0.02 & 0.69 $\pm$ 0.03  \\
(f) WFI~J2026-4536 & 60.8 $\pm$ 5.4 & 0.24 $\pm$ 0.03 & 0.72 $\pm$ 0.03 & 0.57 $\pm$ 0.02 & 0.64 $\pm$ 0.02 \\
(g) HS~2209+1914 & 63.1 $\pm$ 3.25 & 0.05 $\pm$ 0.02 & 0.55 $\pm$ 0.01 & 0.52 $\pm$ 0.01 & 0.53 $\pm$ 0.01 \\
\hline
\end{tabular}
\vspace{0.2cm}
\caption{Measured shape parameters for the lensing galaxy.}
\label{gal7}
\end{table*}

\subsection{Discussion}

A few remarks can be made about the results from the deconvolution:
\begin{itemize}

 \item \textbf{HE~0047-1756 (a)}

A faint Einstein ring, stretched image of the quasar host galaxy, is revealed by the deconvolution. 
\\
 \item \textbf{RX~J1131-1231 (b)}

\citet{Sluse2006} reports astrometric measurements on the same frames with the MCS deconvolution algorithm but with no iterative strategy. Their results agree within the error bars with those presented here. An offset of up to 3 mas between both results is observed. This difference is probably due to the large brightness of the Einstein ring. Indeed, the different amount of recovered background under the PSF can lead to a small shift in position. Photometry is also affected by the presence of the ring. We derive an absolute photometry about 0.4 mag brighter than \citet{Sluse2006} but we obtain compatible flux ratios. The remnant systematic structures in the final residual map also result from the presence of this very bright ring which affects the PSFs and degrades their quality. Indeed, because a part of the background is identical under the three brightest lensed images (and thus with more weight in the determination of the PSF), it is impossible to completely disentangle the flux contribution of the ring from the one really included in the point sources. This is a limitation of the ISMCS method: for it to work properly, the background has to be different under each point source.
\\
 \item \textbf{SDSS~J1138+0314 (c)}

A faint Einstein ring is revealed by the deconvolution process. 
\\
 \item \textbf{SDSS~J1155+6346 (d)}

Our astrometry is not in agreement with \citet{Pindor2004} especially concerning the lens. The difference for source B amounts to 0\farcs1 in RA and 0\farcs04 in DEC while for the lens the offset is much larger and amounts to 1\farcs55 in RA and 0\farcs28 in DEC. However, our astrometry is in agreement with the one listed in the CASTLES database. The remnant systematic structures in the final residual map are due to the presence of the very bright and extended lensing galaxy (which is not clear on the presented frames, the cuts being chosen so that the two lensed images appear clearly). As in the case of RX~J1131-1231, it is not possible to completely disentangle the background flux contribution from the one of the point sources, a part of the background being identical under both lensed images. This degrades the quality of the PSFs and of the deconvolution.
\\

\item \textbf{WFI~J2026-4536 (f)}

The astrometry we obtain, except for the right ascension of the lens ($\Delta RA=$0\farcs03), is in agreement, within the error bars, with the results of \citet{Morgan2004} who used the same frames but a different image processing technique.
\\
 \item \textbf{HS~2209+1914 (g)}

A bulge is clearly observable but an additional extended structure is also visible. It could either be some spiral arms, in which case the lens would be a late-type galaxy, or even a distorted Einstein ring. A spectrum of the lens and higher resolution imaging would help to disentangle between these hypotheses. As we do not know what this structure is related to, we fit a de Vaucouleurs model on the bulge only, in using a special feature of the MCS algorithm: a mask encircling the lens galaxy to avoid the model to fit this extended structure. Moreover, the residual map contains many intense structures. However, as these structures do not have the same shape under both point sources, we cannot recover them with another iteration of ISMCS. Saturation is unlikely to be responsible for that phenomenon, as it is corrected by the NICMOS reduction pipeline. It could thus be due to differential extinction by the lensing galaxy, resulting in a different color for both lensed images and thus a different shape of the PSF.

\end{itemize}
Let us note that all our results are in agreement with what can be found in the CASTLES database, within their error bars (ours being smaller).

\begin{figure*} [h!]
\centering
  \subfigure{\includegraphics[scale=0.7]{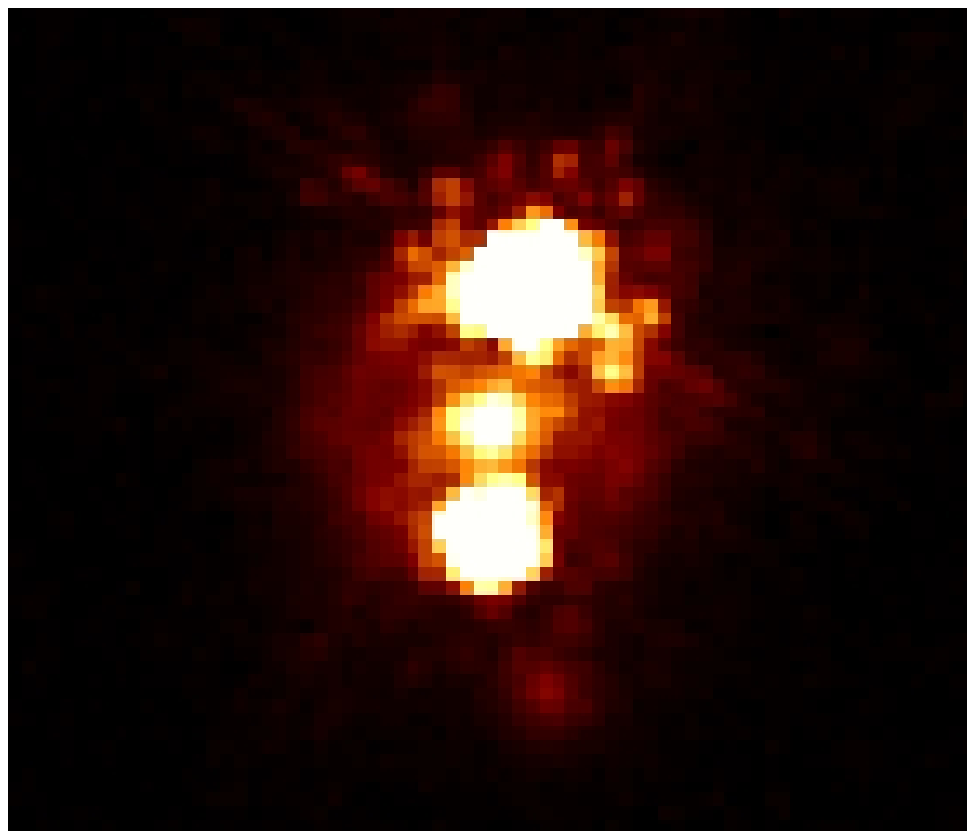}}                
  \subfigure{\includegraphics[scale=0.565]{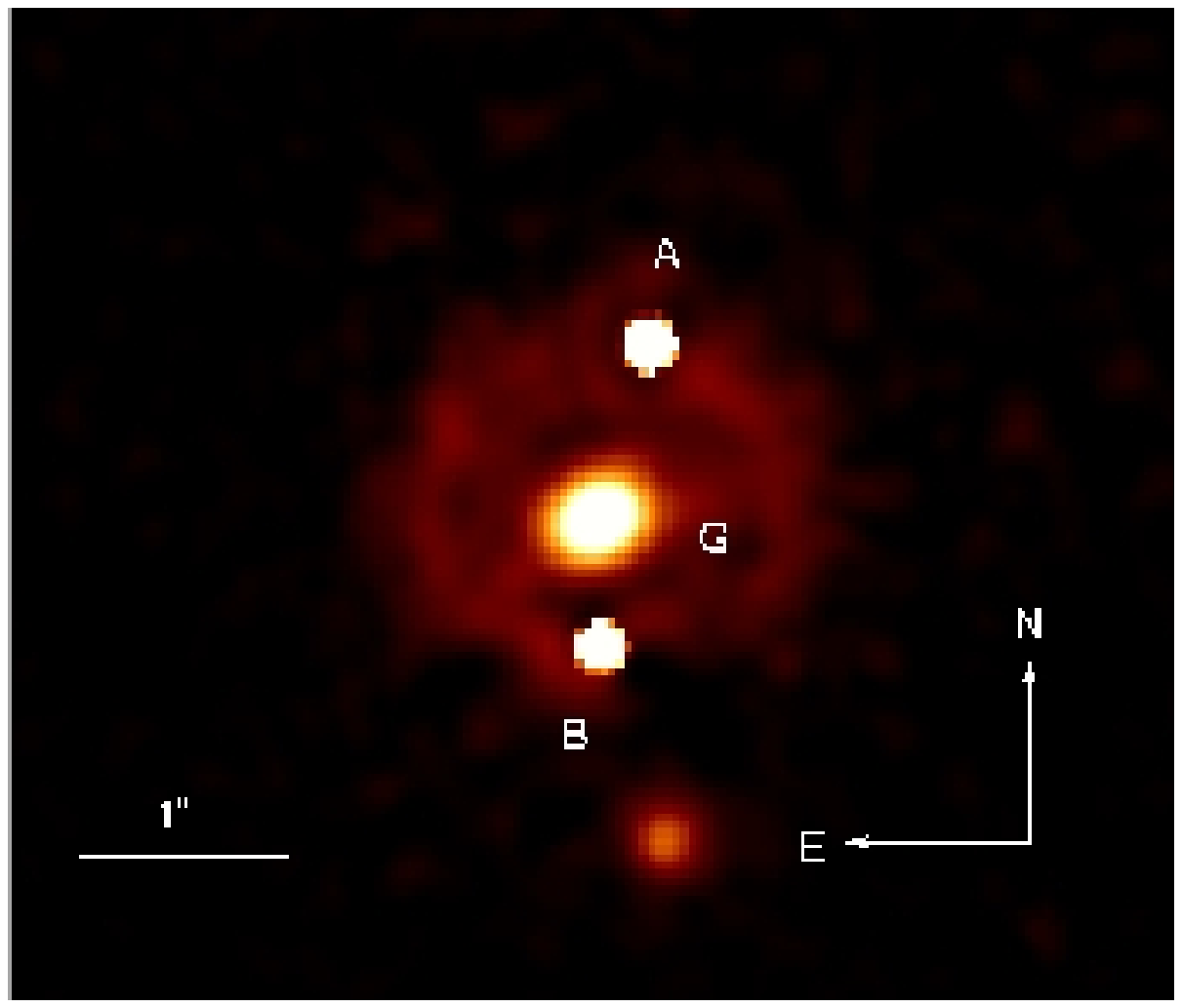}}
\setcounter{subfigure}{0}
  \subfigure[a][HE~0047-1756]{\includegraphics[scale=0.7]{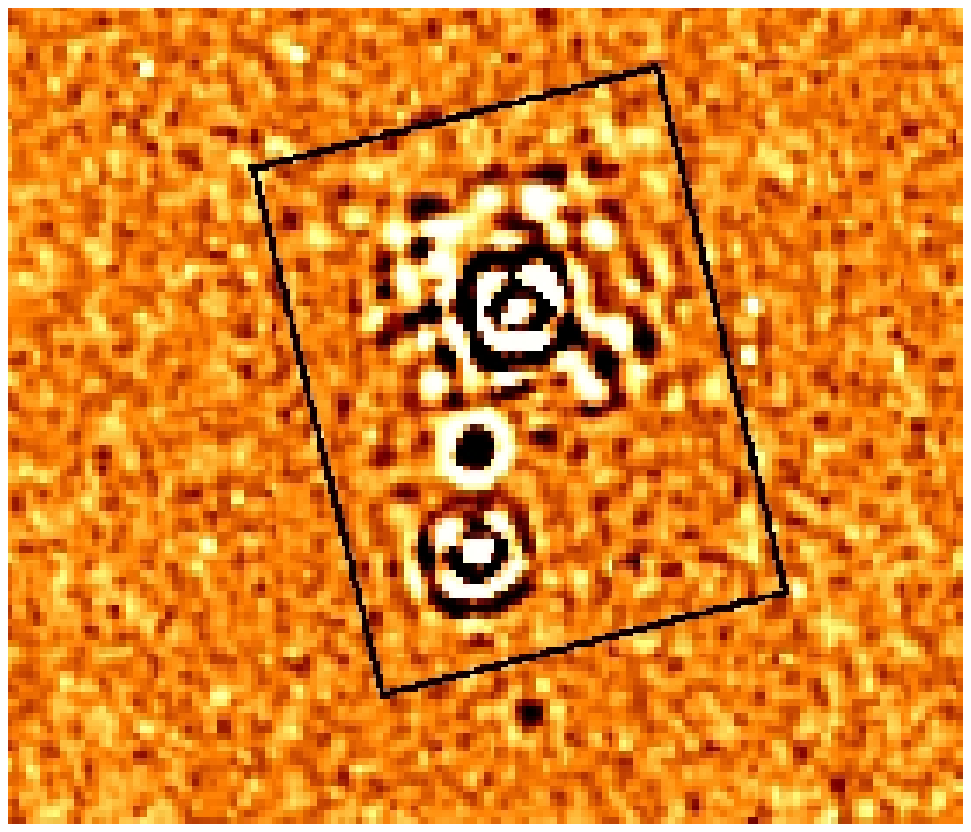}}
  \subfigure{\includegraphics[scale=0.7]{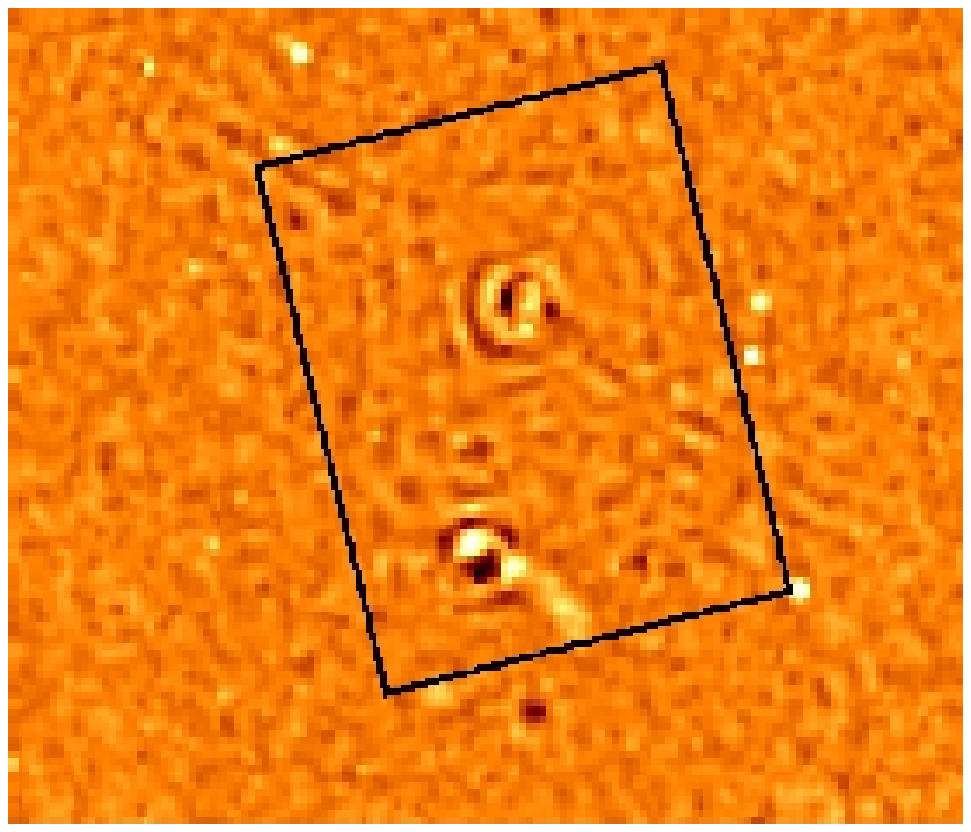}}
  \subfigure{\includegraphics[scale=0.7]{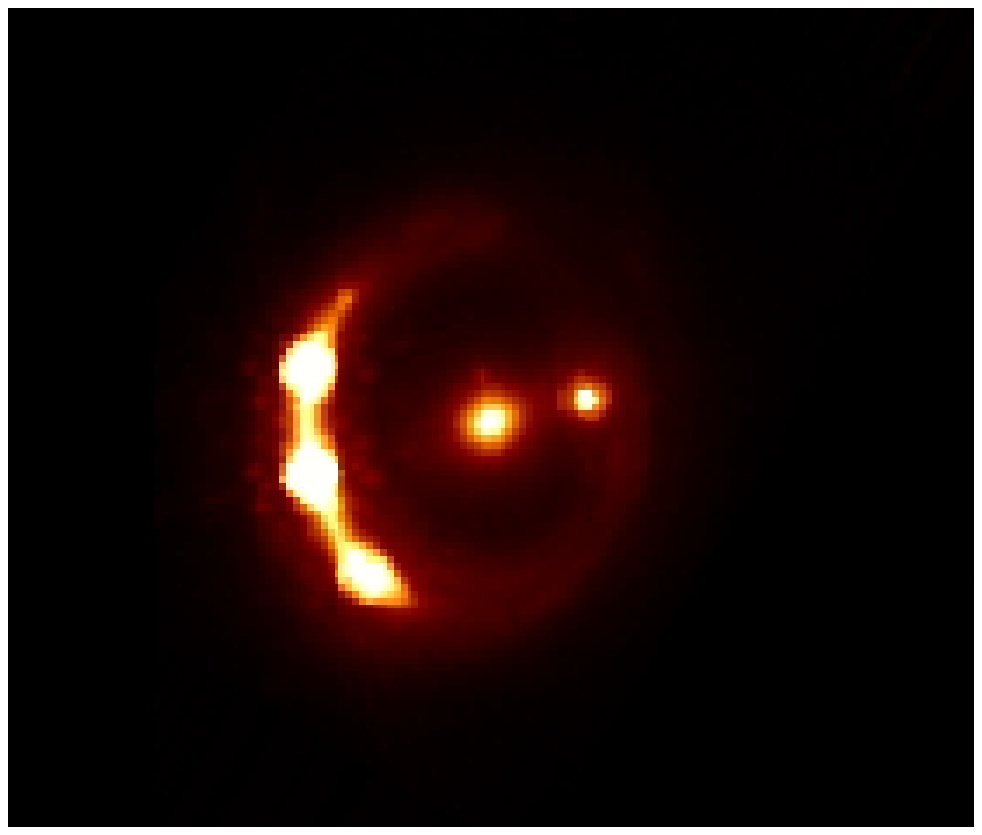}}                
  \subfigure{\includegraphics[scale=0.566]{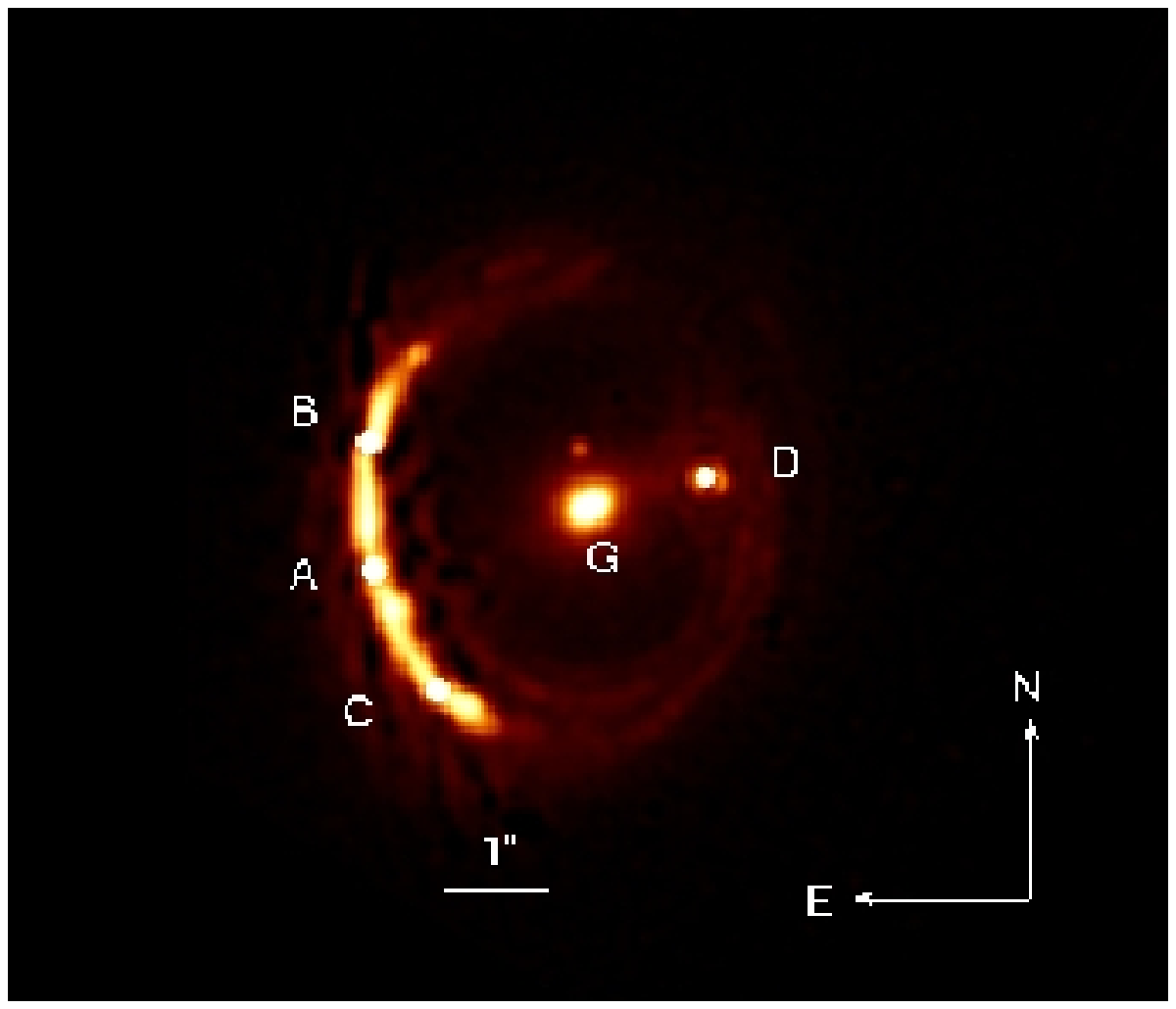}}
\setcounter{subfigure}{1}
  \subfigure[RX~J1131-1231]{\includegraphics[scale=0.725]{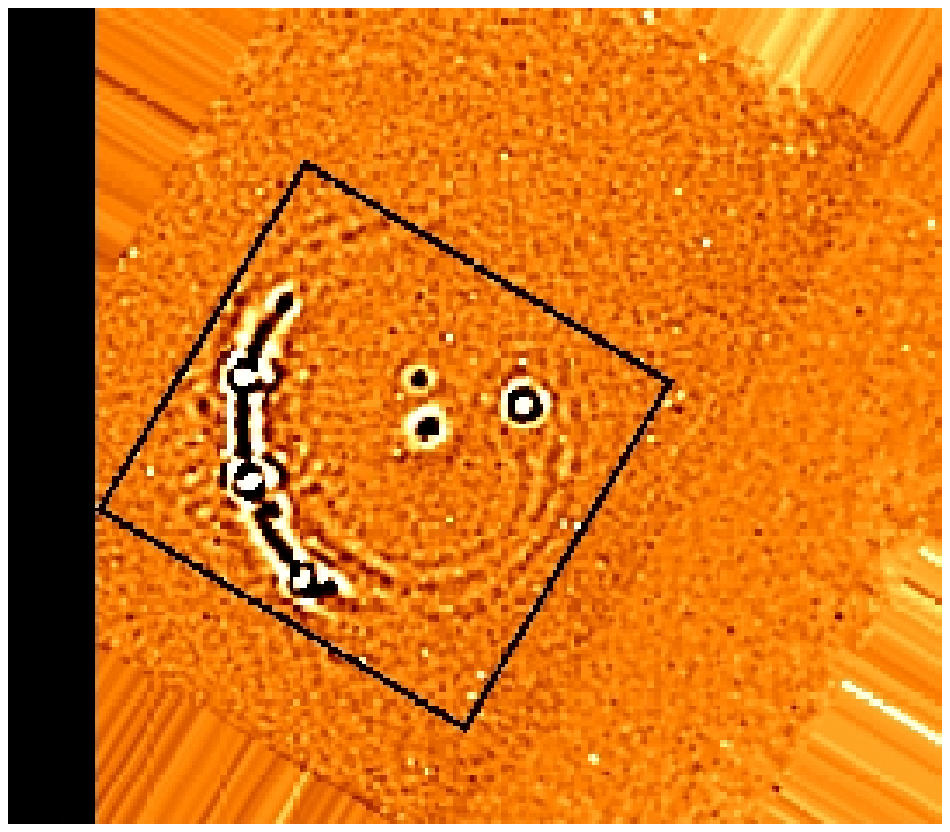}}
  \subfigure{\includegraphics[scale=0.725]{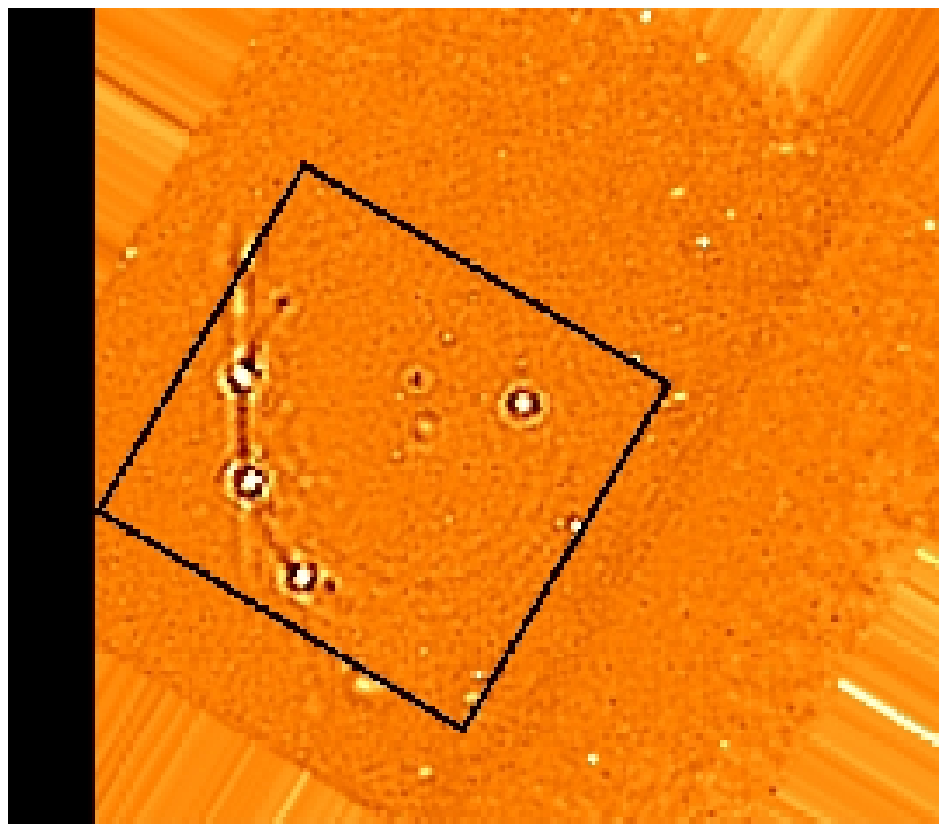}}
\caption{HST/NIC2 original and deconvolved frames (resp. top left and top right), mean residual map from the first and from the last iteration of ISMCS (resp. bottom left and bottom right).}
\label{dec_NICMOS7}
\end{figure*}

\begin{figure*} [h!]
\centering  \subfigure{\includegraphics[scale=0.7]{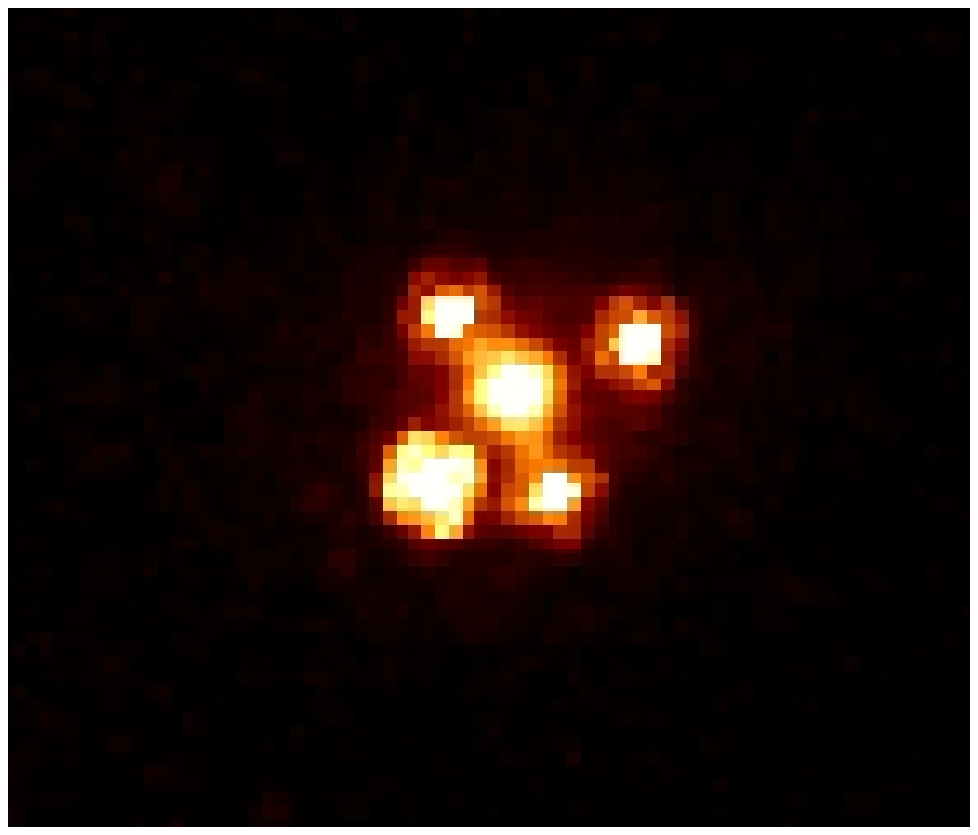}}                
  \subfigure{\includegraphics[scale=0.565]{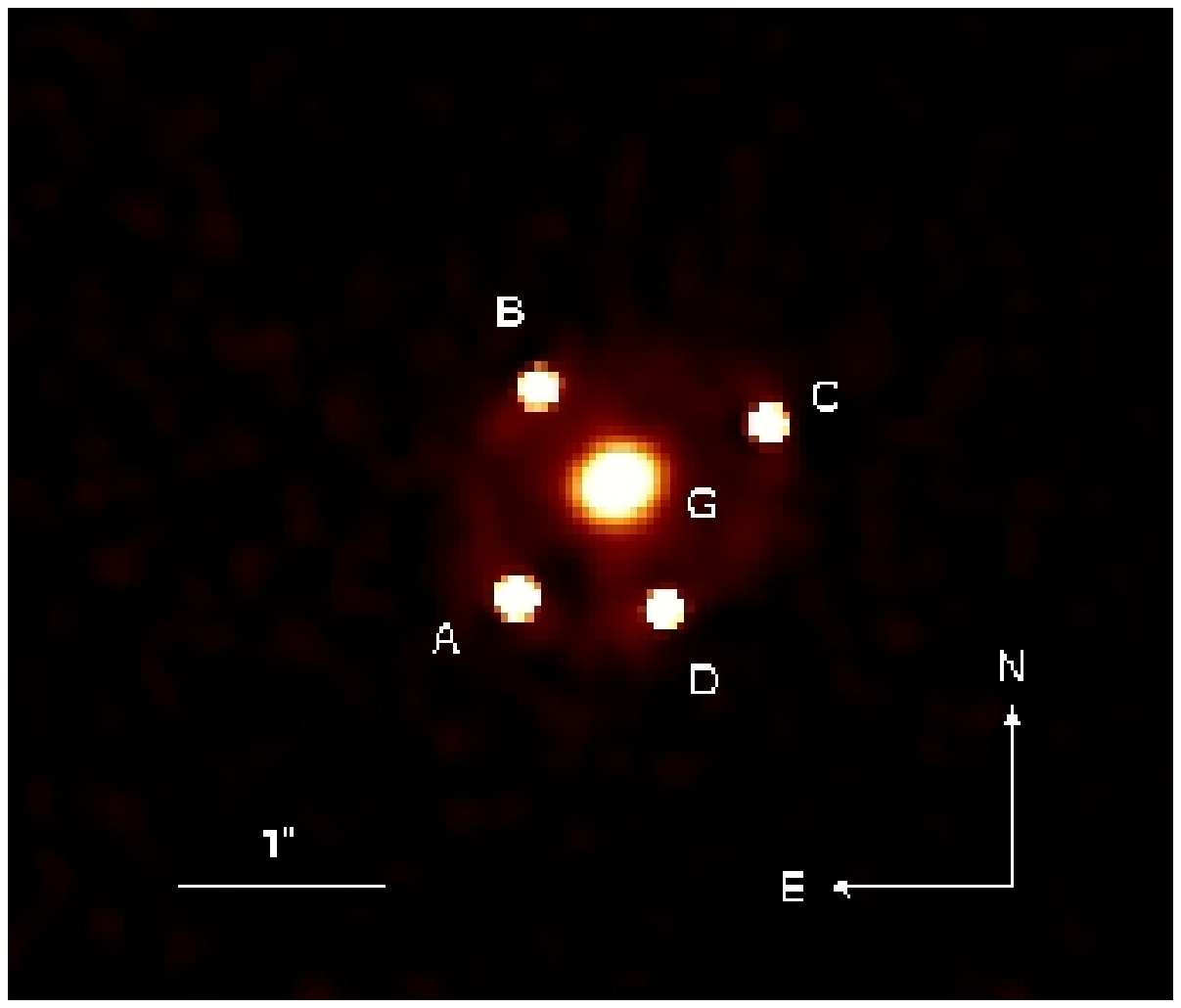}}
\setcounter{subfigure}{2}
  \subfigure[SDSS~J1138+0314]{\includegraphics[scale=0.7]{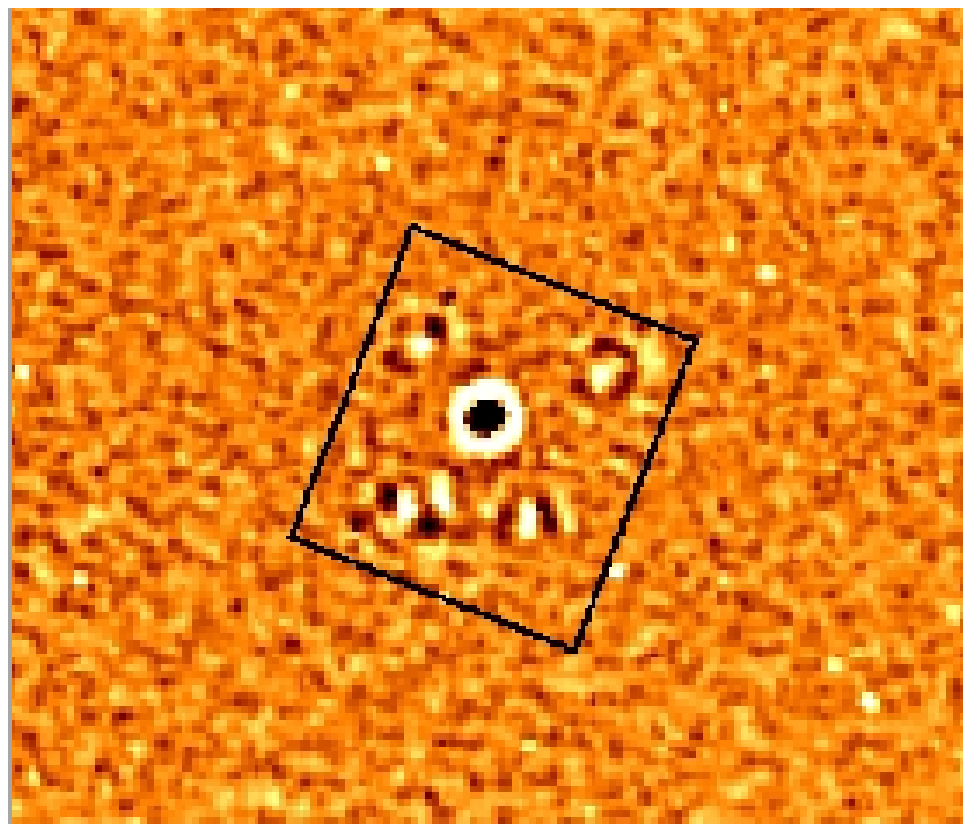}}
  \subfigure{\includegraphics[scale=0.7]{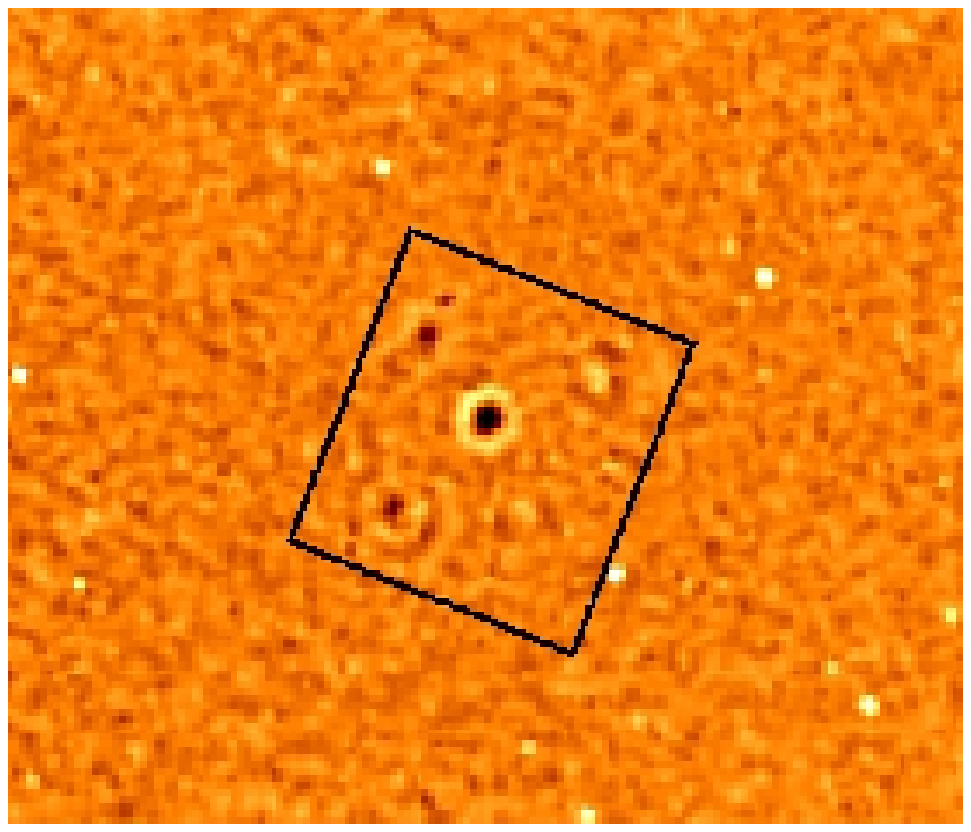}}
  \subfigure{\includegraphics[scale=0.7]{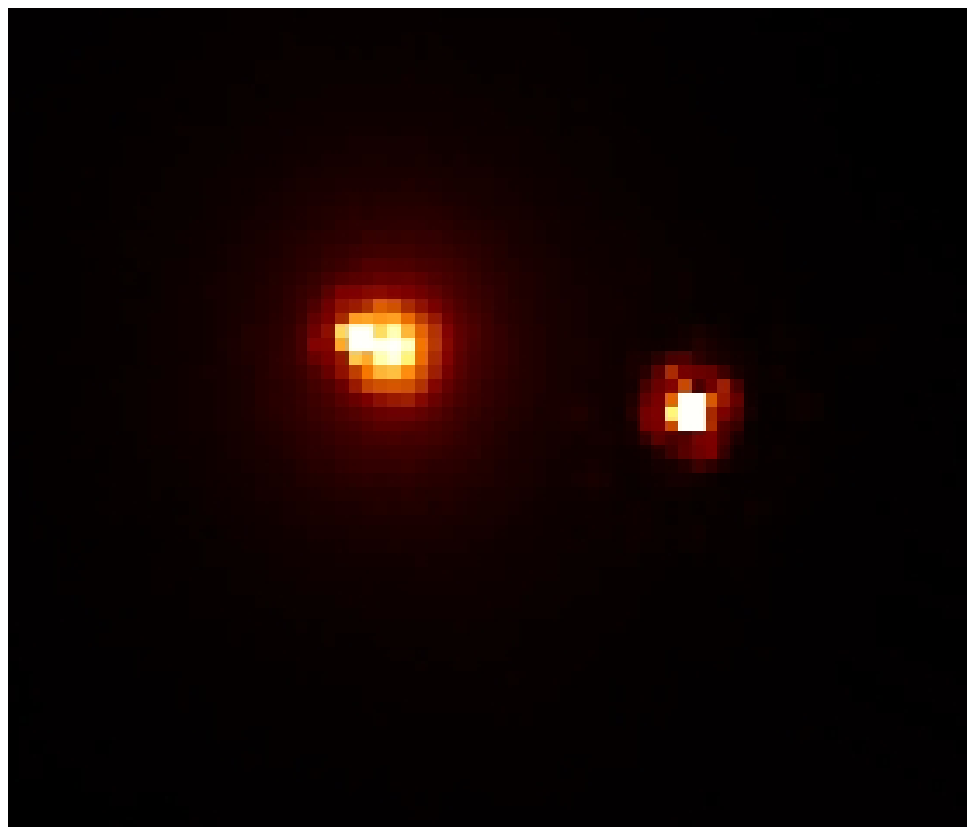}}                
  \subfigure{\includegraphics[scale=0.565]{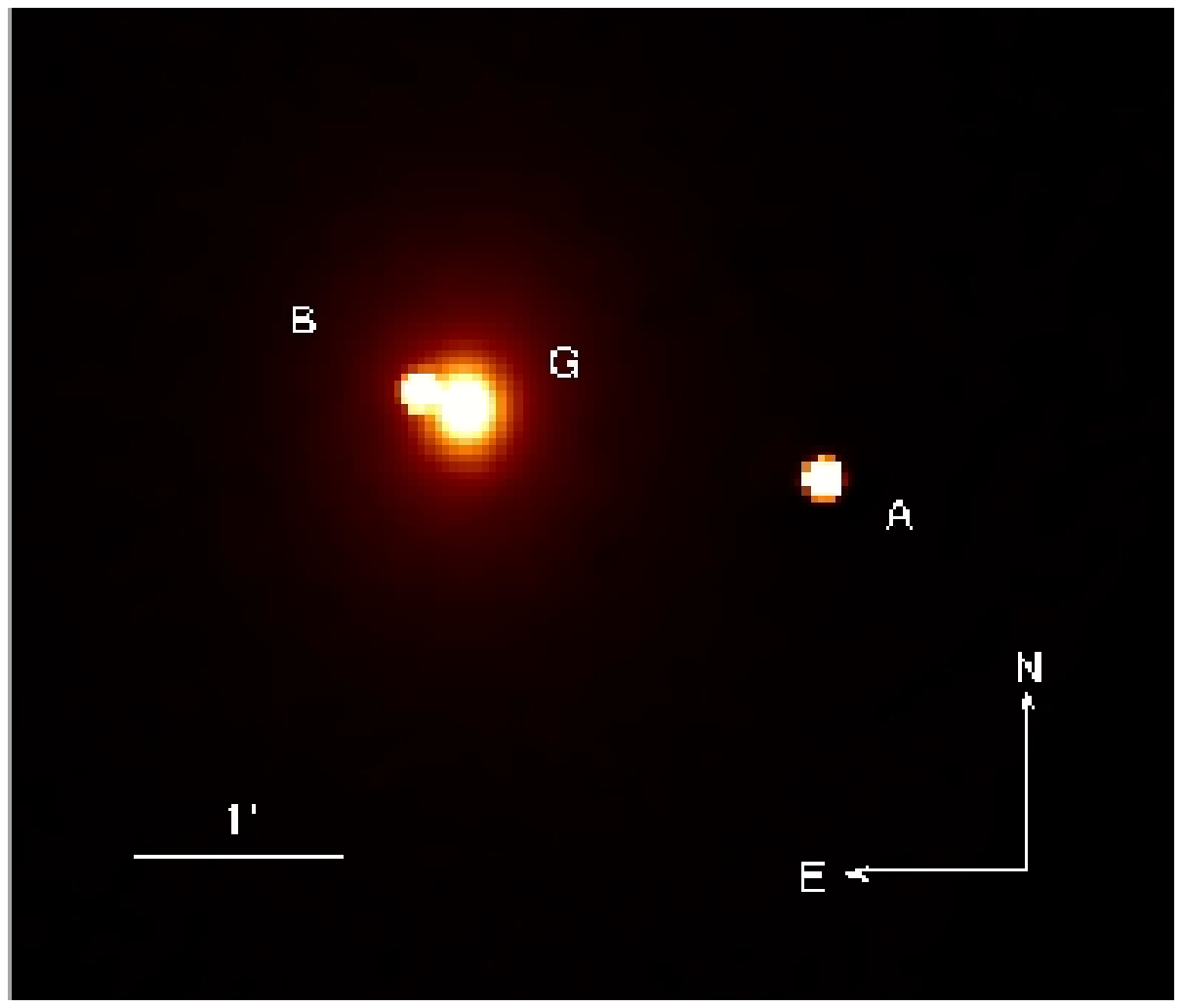}}
\setcounter{subfigure}{3}
  \subfigure[SDSS~J1155+6346]{\includegraphics[scale=0.7]{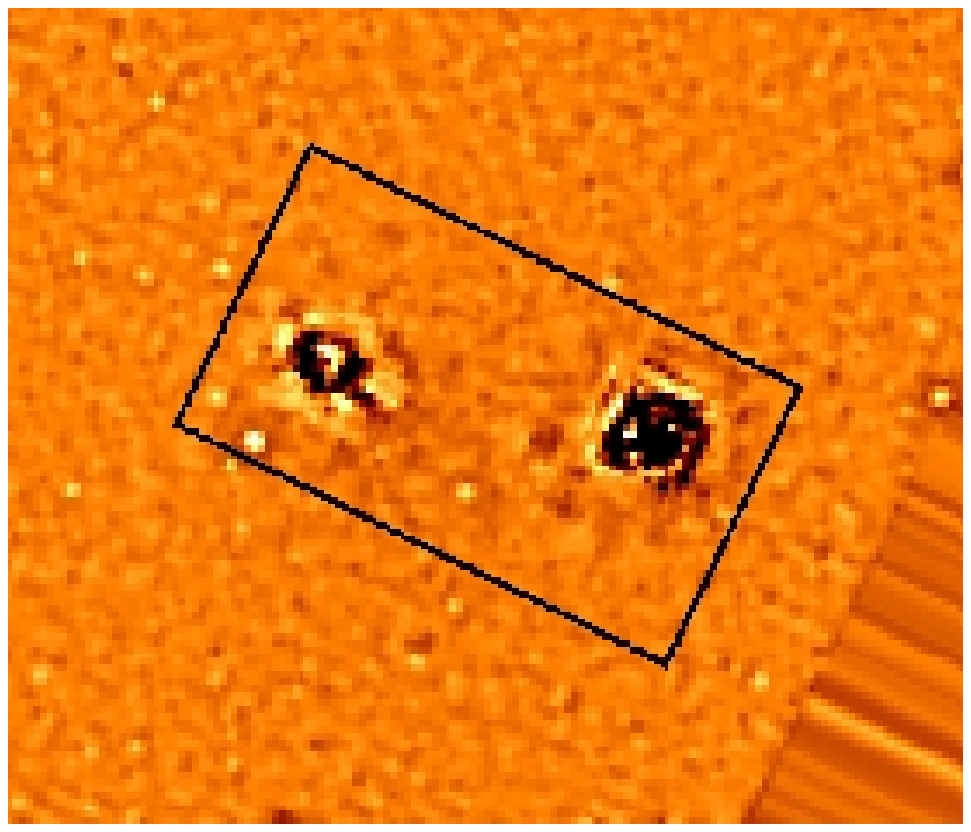}}
  \subfigure{\includegraphics[scale=0.7]{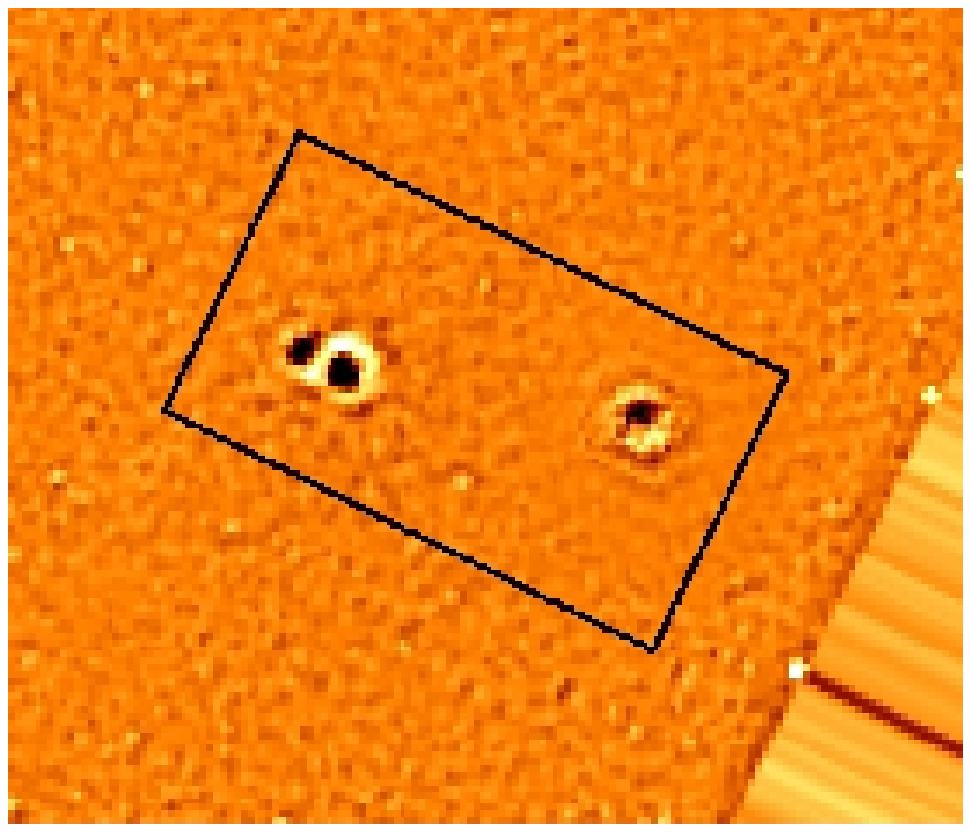}} 
\setcounter{figure}{0}
\caption[]{continued.}
\end{figure*}

\begin{figure*} [h!]
\centering
  \subfigure{\includegraphics[scale=0.7]{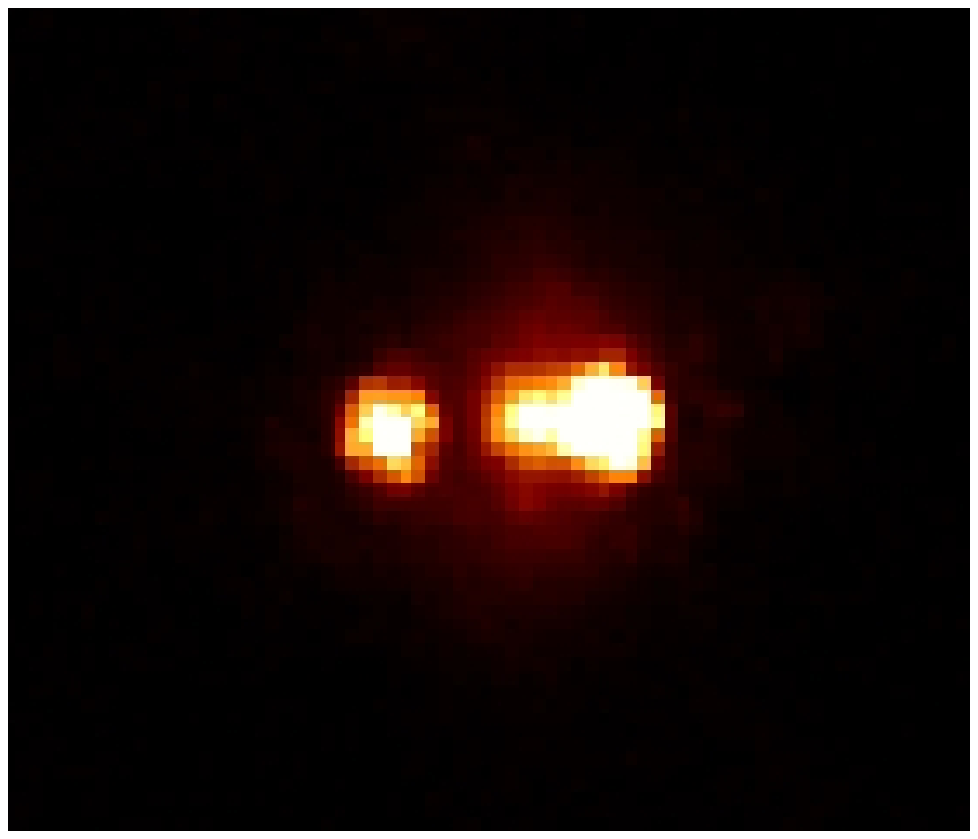}}                
  \subfigure{\includegraphics[scale=0.565]{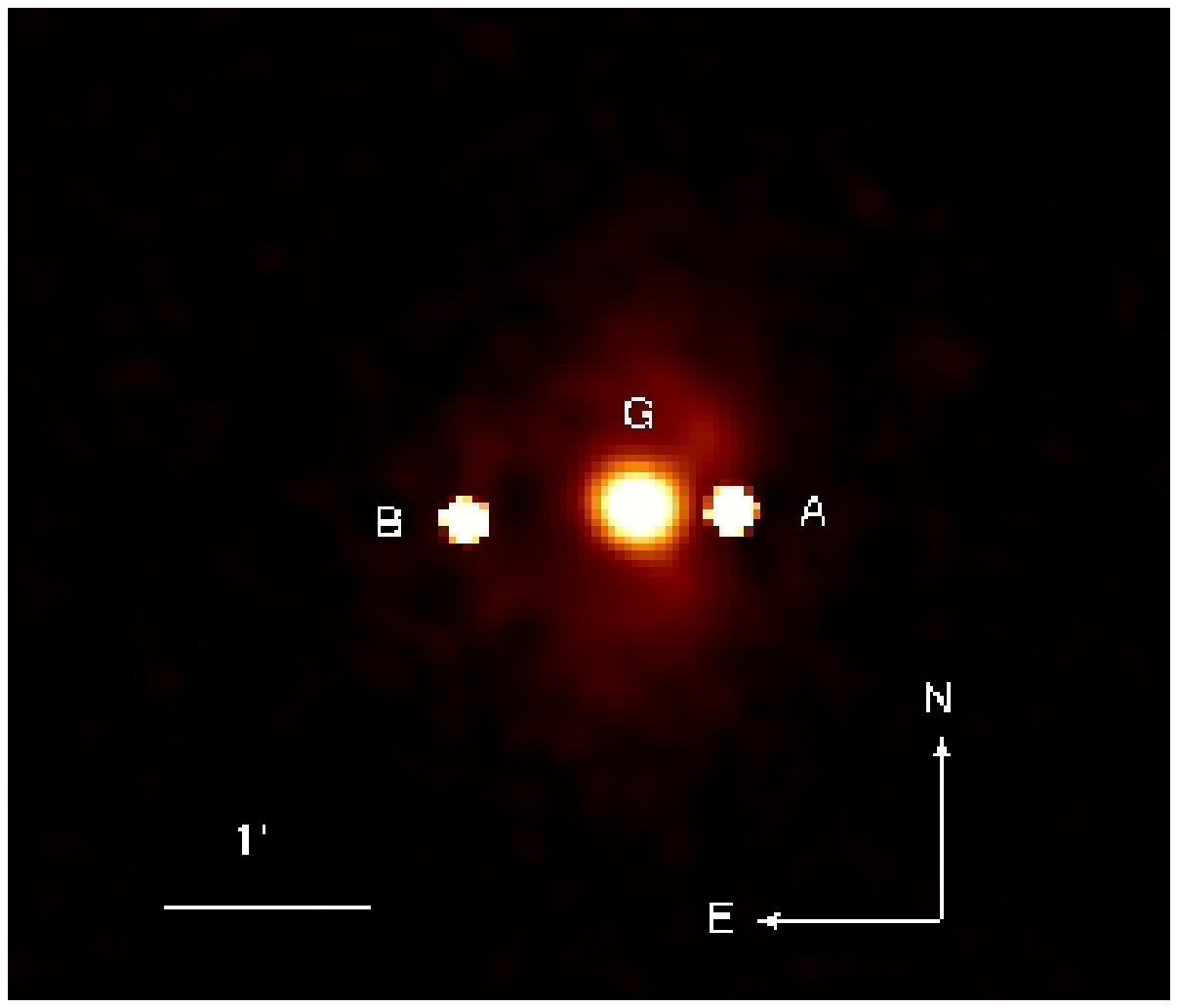}}
\setcounter{subfigure}{4}
  \subfigure[SDSS~J1226-0006]{\includegraphics[scale=0.7]{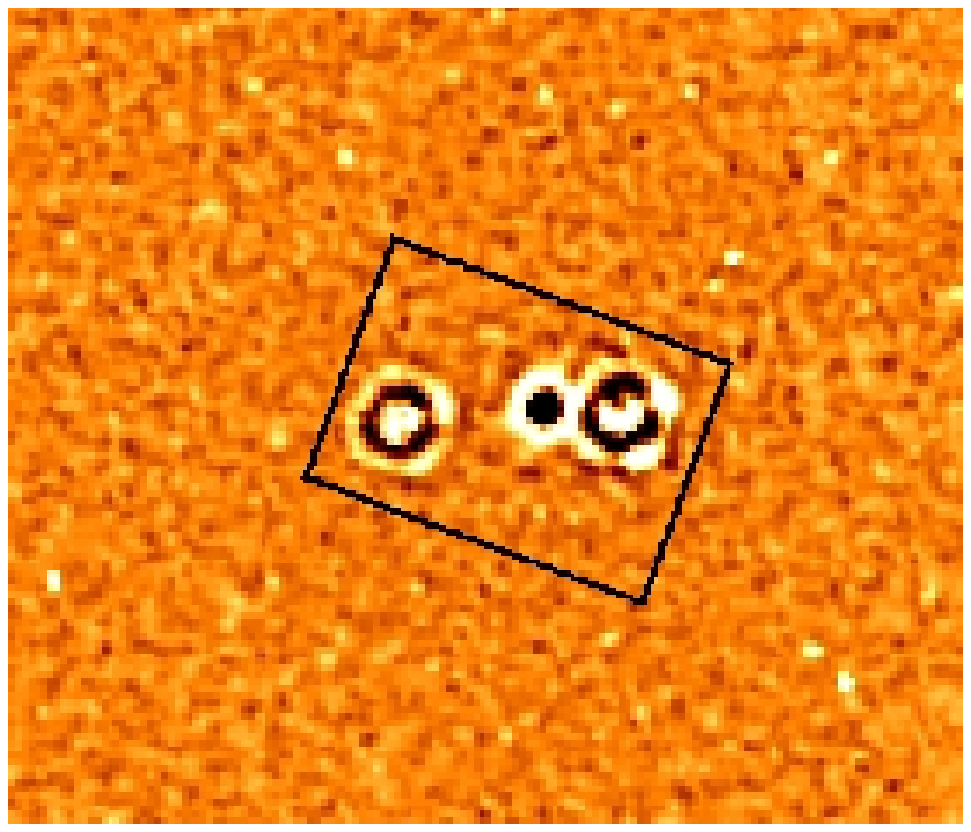}}
  \subfigure{\includegraphics[scale=0.7]{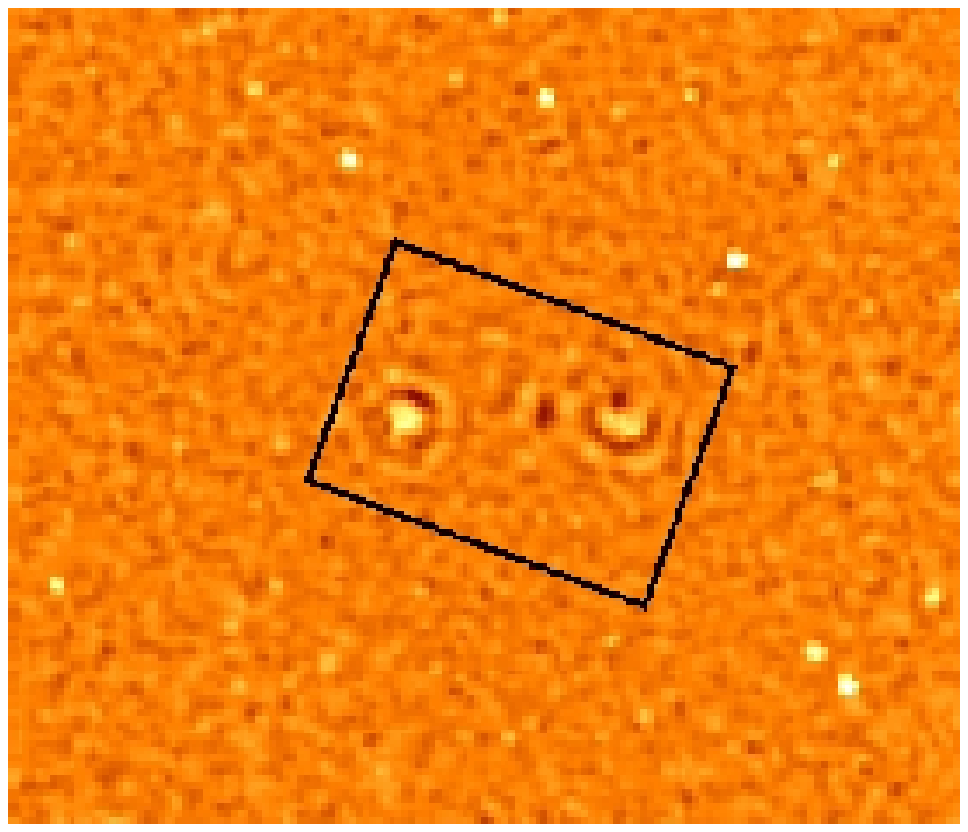}}
  \subfigure{\includegraphics[scale=0.7]{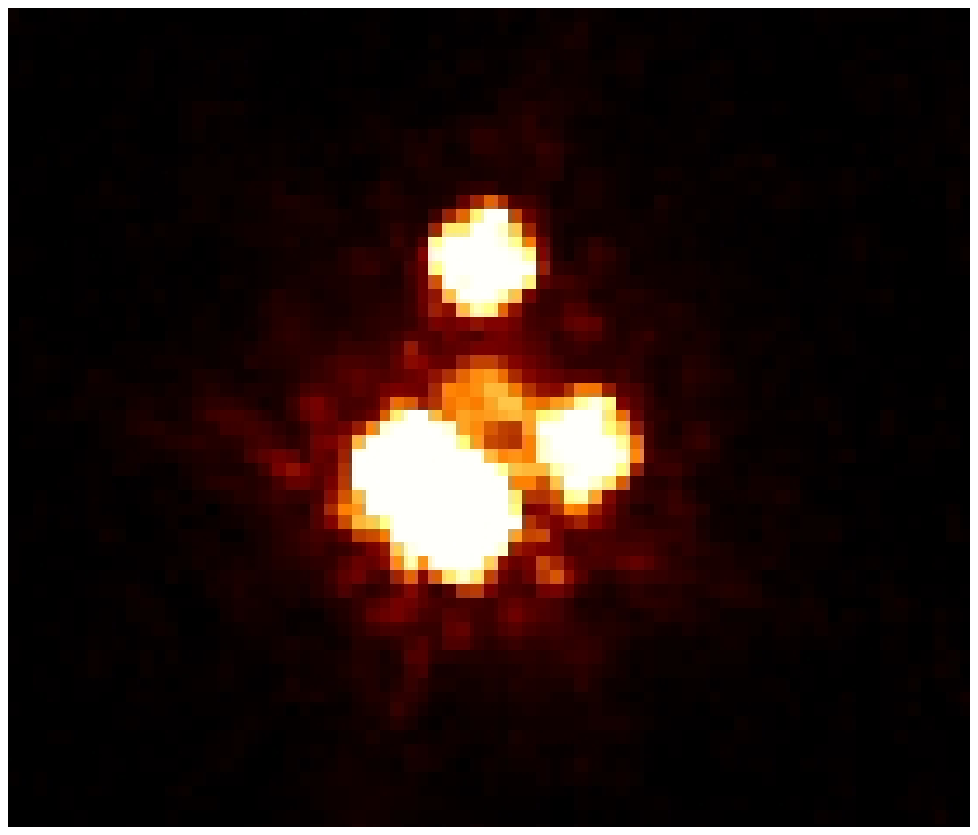}}                
  \subfigure{\includegraphics[scale=0.563]{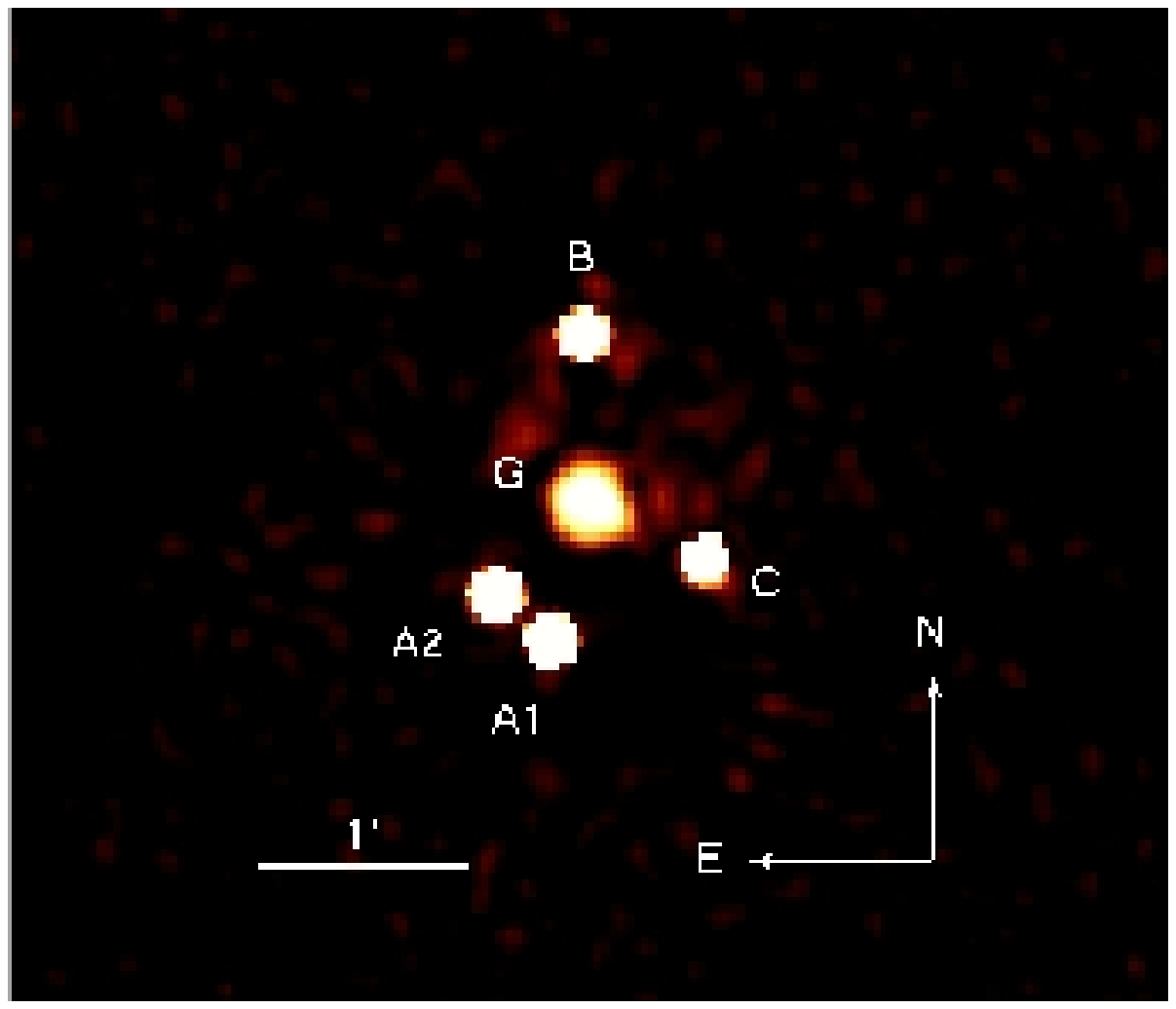}}
\setcounter{subfigure}{5}
  \subfigure[WFI~J2026-4536]{\includegraphics[scale=0.7]{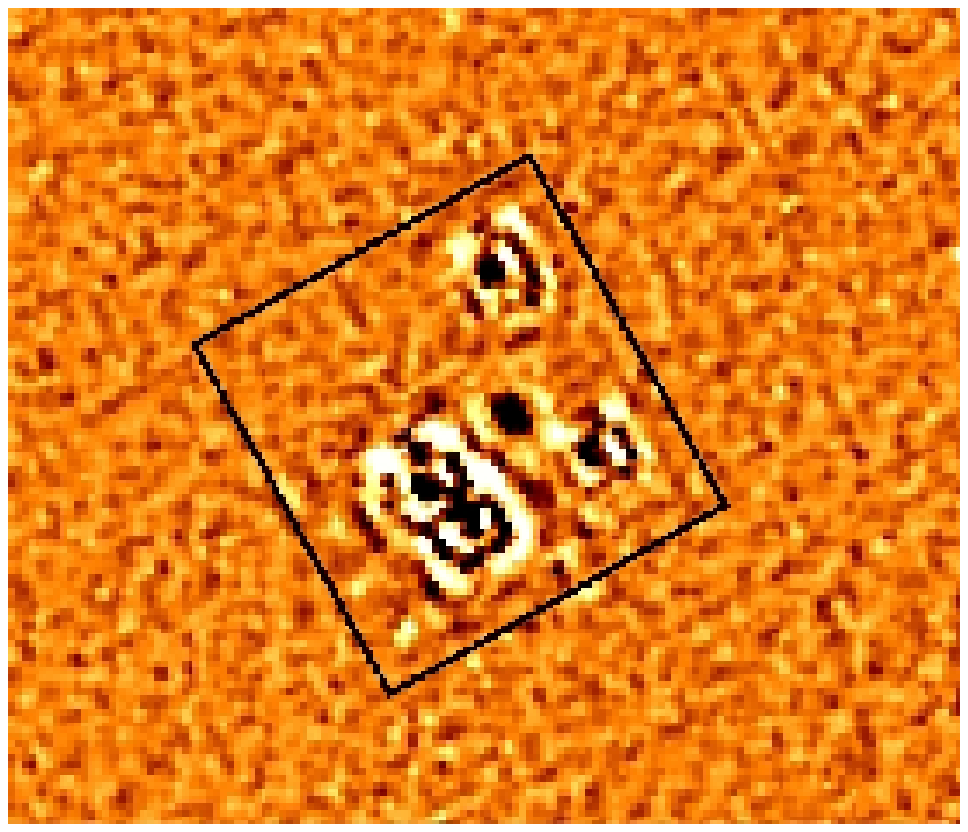}}
  \subfigure{\includegraphics[scale=0.7]{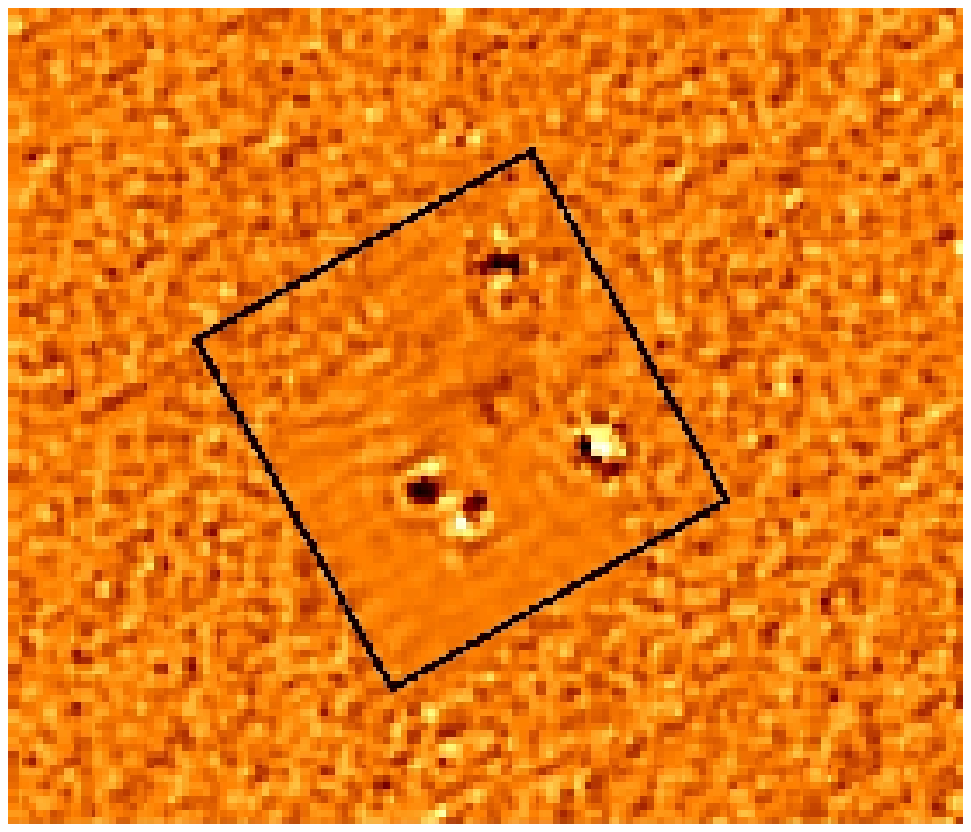}}
\setcounter{figure}{0}
\caption[]{continued.}
\end{figure*}

\begin{figure*} [h!]
\centering
  \subfigure{\includegraphics[scale=0.7]{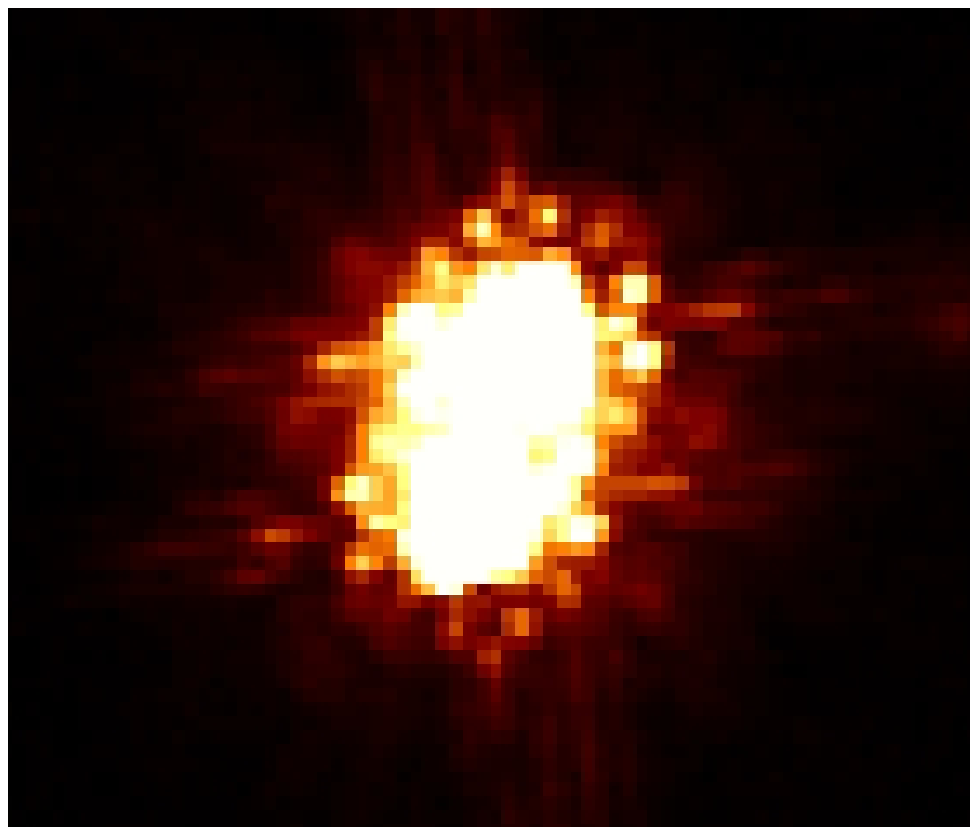}}                
  \subfigure{\includegraphics[scale=0.565]{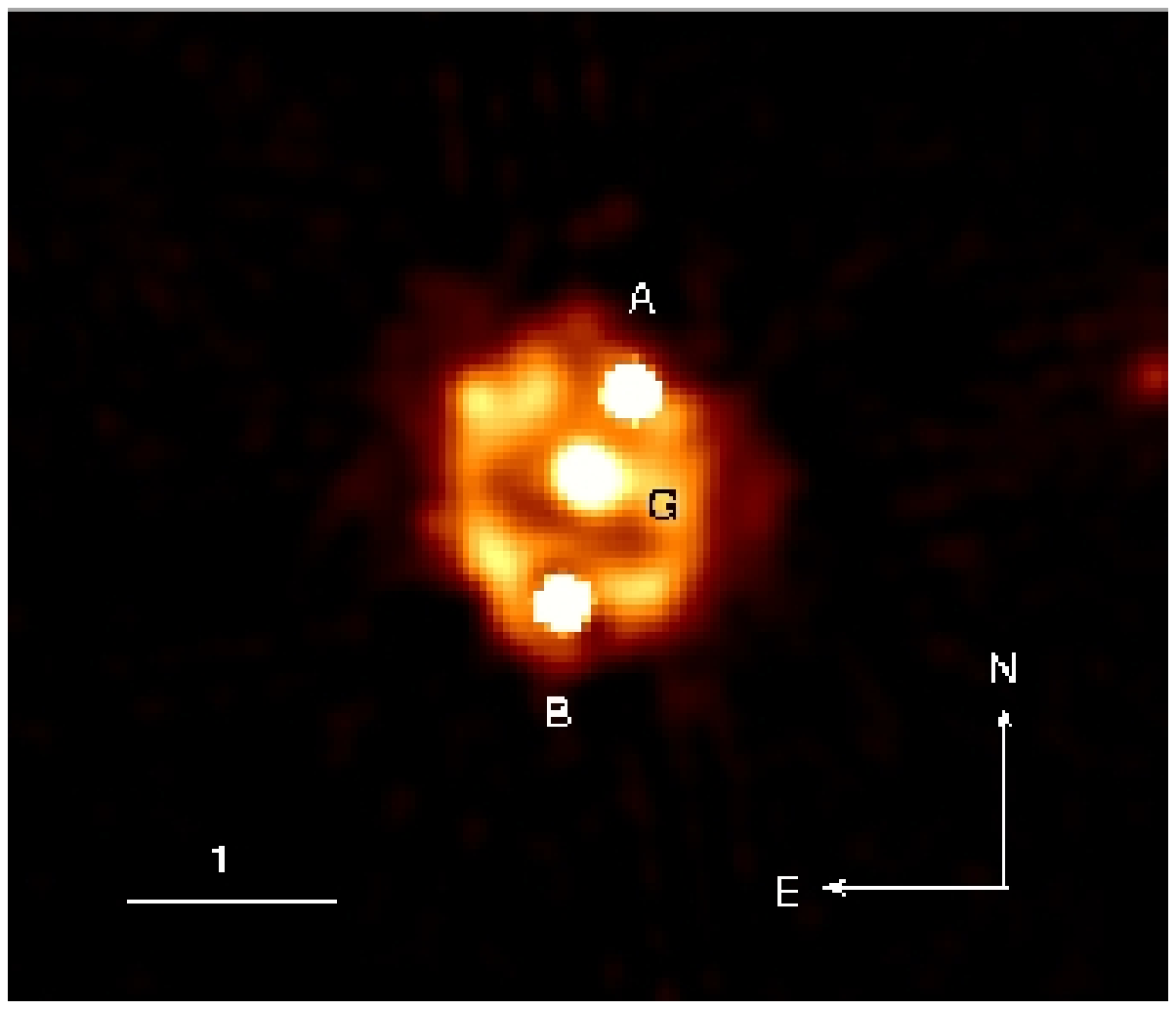}}
\setcounter{subfigure}{6}
  \subfigure[HS~2209+1914]{\includegraphics[scale=0.72]{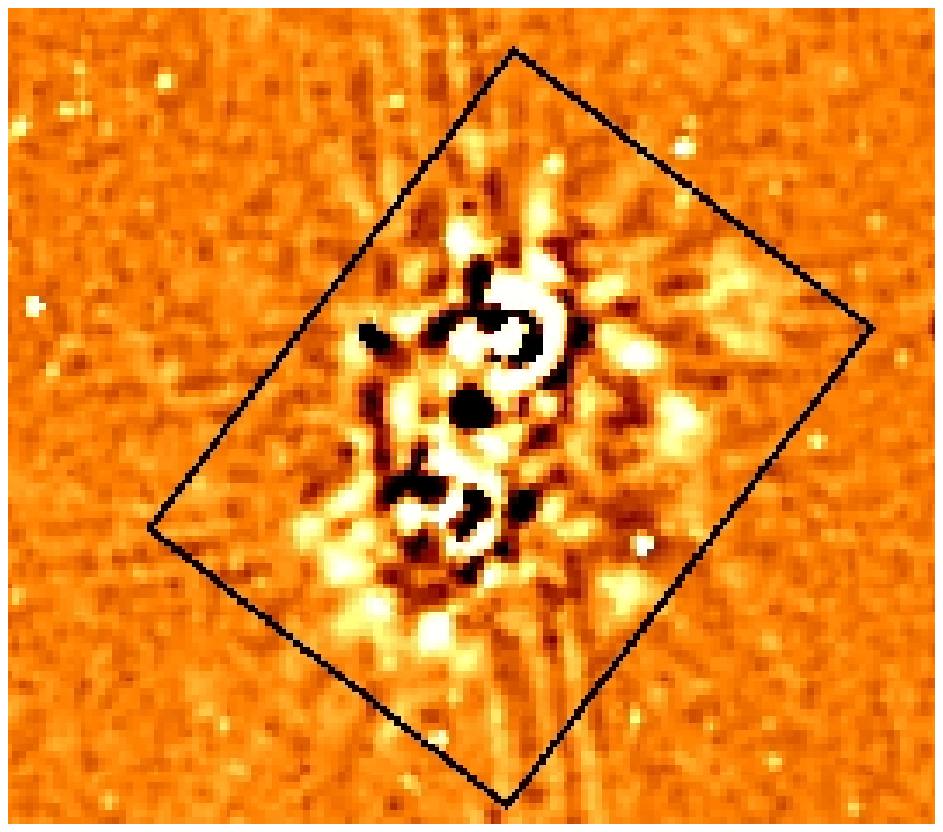}}
  \subfigure{\includegraphics[scale=0.72]{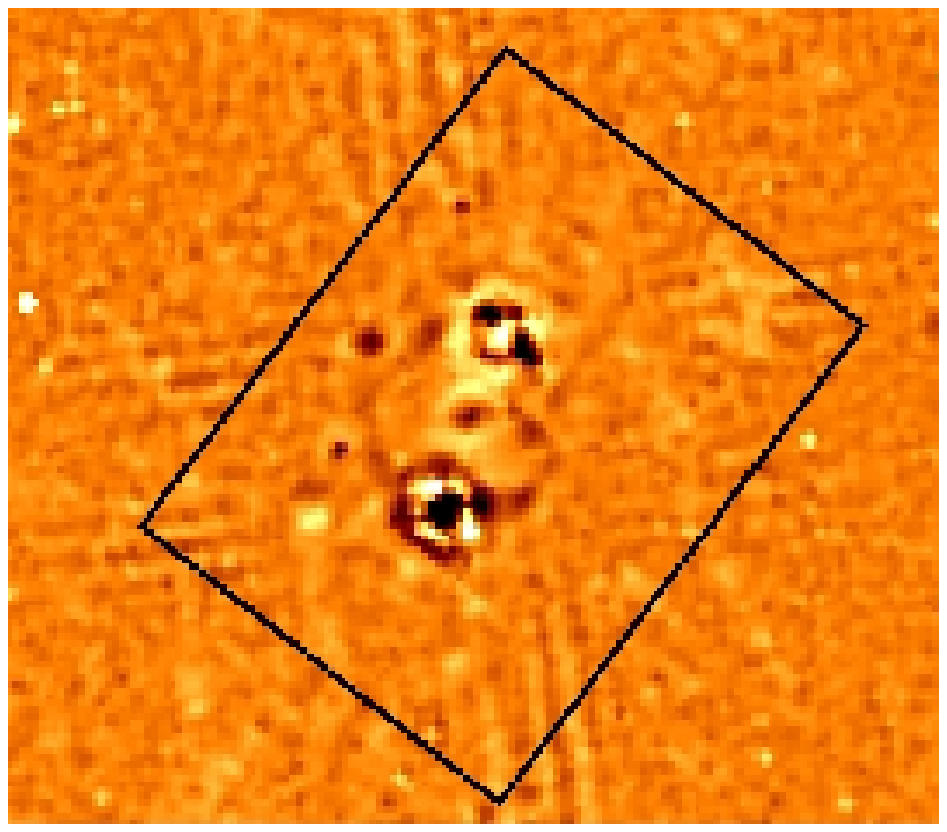}}
\setcounter{figure}{0}
\caption[]{continued.}
\end{figure*}

\section{Parametric modeling}
\label{model7}
Our goals are twofold. First, we aim at providing simple models and prospective time delays for the lensed quasars monitored by COSMOGRAIL. Second, we aim to test whether simple smooth lens models are able to reproduce the mas relative astrometry of quadruply imaged quasars in 3 systems without measured time delays. Our strategy consists in using the LENSMODEL software package v1.99o \citep{Keeton2001} to model the mass distribution of our seven systems. For a chosen model of the mass distribution, the code minimizes a $\chi^{2}$ defined as the square of the difference between observable quantities and their ``model counterparts'', weighted by the observational errors on these quantities. Two different lens models are considered. First, an isothermal profile, which is the standard mass distribution to model gravitational lenses \citep{Kassiola1993}, and second, a de Vaucouleurs profile, for which we assume that the light perfectly traces the mass in the inner regions of lensing galaxies. These two models should provide a good approximation of the extreme slopes of the mass distribution at the location of the lensed images and of the expected time delays \citep{Kochanek2002,Kochanek2004}. In addition, the study of the galaxy-galaxy lensing sample from the \emph{Sloan Lens ACS}\footnote{ACS stands for \emph{Advanced Camera for Surveys}.} \emph{Survey} \citep[SLACS,][]{Bolton2006} has revealed that the massive elliptical lensing galaxies are nearly kinematically undistinguishable from isothermal ellipsoids \citep[see e.g.][]{Koopmans2009}. This supports the use of an isothermal gravitational potential as a fiducial model to test the ability of smooth lens models to reproduce quadruply imaged quasars with mas accuracy. Since lensing galaxies are never isolated, we model the effect of the environment with an external shear term characterized by an amplitude $\gamma$ and a position angle $\theta_{\gamma}$ (pointing towards the mass at the origin of the shear). All the models are computed for a flat universe with the following cosmological parameters: $\rm H_{0}=70$ km/s/Mpc, $\Omega_{m}=0.3$ and $\Omega_{\Lambda}=0.7$. 

To model our systems, we use every constraint at our disposal: the relative astrometry of the lensed images, with the MTE, i.e. the uncertainties displayed in the fifth column of Table \ref{astrom7}, the position of the main lens, with the error inherent to the deconvolution when it is larger than the MTE, and, in the case of doubly imaged quasars, the flux ratio between the two point sources. In principle, fluxes can be contaminated by different effects such as microlensing by stars in the lens galaxy, dust extinction and also by the time delay itself. As the flux ratios are measured in the near-infrared, all these effects should be small \citep{Yonehara2008}. We thus assume a 1$\sigma$ error of 10\% on the flux ratios. In summary, we have 10 constraints for the quads, while we have 8 for the doubles. For the de Vaucouleurs model, we assume that the total mass profile follows the light profile. We thus add three constraints to the model: the PA of the galaxy, its ellipticity $e$ and effective radius $R_{eff}$ (see Table \ref{gal7}). Due to the limited number of observational constraints, isothermal mass profiles are assumed to be spherically symmetric (SIS, i.e. Singular Isothermal Sphere) when modeling doubles. This is not a strong asumption as the quadrupole term of the potential modifies only slightly the time delays of doubly imaged quasars \citep{Kochanek2002, Wucknitz2002}. For quads, we allow the ellipticity of the isothermal mass distribution (SIE, i.e. Singular Isothermal Ellipsoid) to deviate from the ellipticity of the light profile. This enables to account for dark matter halos that would be rounder then the light distribution \citep{Ferreras2008}. The position angle of the total mass distribution can be constrained  as the one of the light profile as these two distributions might only be slightly misaligned \citep{Keeton1997a, Ferreras2008}.  Finally, we also assume that the center of the total mass distribution and the one of the light profile are identical within the error bars. This is supported by the work of \citet{Yoo2006} who found, for 4 lensed quasars with an Einstein ring, that the offset between the light and the total mass distribution is limited to a few mas. Calculating the number of degree(s) of freedom (d.o.f.), which is the difference between the number of model parameters and observable quantities, we find 0 d.o.f. when modeling doubly imaged quasars and 2 (resp. 3) d.o.f. when modeling quads with SIE (resp. de Vaucouleurs) + external shear.

The search for the best model and estimate of uncertainties is performed in two steps. First, we generate an initial sample of 2000 different models with parameters distributed over the whole parameter space and optimize them. This method is efficient to find the best models and identify local minima. Then, in order to estimate the model uncertainties, we sample the posterior probability distribution of the parameter space using an adaptive Metropolis Hastings Monte-Carlo Markov Chain (MCMC) algorithm. This technique is implemented in LENSMODEL and described in \citet{Fadely2010}. In practice, an ensemble of 15 different chains are run, each chain consisting of a sequence of trial steps drawn from a multivariate gaussian distribution of width estimated thanks to the first step of the process. The sampling of the parameter space is optimised by using the covariance matrix. In 5\% of the steps, the covariance matrix is diagonal, allowing to use a large step along one of the axis and better escape local minima in the $\chi^2$ surface. We use the same criterion as \citet{Fadely2010} to assess that any MCMC run has converged. Finally, for each point of the MCMC, we calculate the relative likelihood of a parameter $p$ based on the $\chi^2$ statistics (i.e. $L(D|p) = exp(-\chi^2/2)$), and calculate a 68\% confidence interval for each parameter.

The parameters of the best fit models are displayed in Table \ref{lensmodel7}. The columns display the following items: the name of the object, the type of mass distribution used (``DV'' stands for \textit{de Vaucouleurs profile}), the mass scale parameter (the angular Einstein radius $R_{Ein}$ in arcseconds), the mass distribution ellipticity $e$ and its orientation $\theta_{e}$ in degrees positive East of North, the effective radius $R_{eff}$ in arcseconds in the case of a de Vaucouleurs model, the intensity of the shear $\gamma$ and its orientation $\theta_{\gamma}$ in degrees (East of North), the number of degree(s) of freedom (d.o.f.), the $\chi^{2}$ of the fit and the predicted time delays in days when the lens redshift is known. For the quads and in the same column as the $\chi^{2}$, we also give the $\chi^{2}_{im}$ which is the contribution of the lensed images position to the $\chi^{2}$, and $\chi^{2}_{l}$ which is the contribution of the lens galaxy position to the $\chi^{2}$. Let us note that $\Delta t_{AB} >$ 0 means that the flux of A varies before the one of B. The median value of each parameter along with 68\% confidence level is shown in Table \ref{MCMC7}.

\begin{table*}[t!]
\centering 
\begin{tabular}{c||c|cccccccc}
\hline
Object & Model & $R_{Ein}$ & $e, \theta_{e}$ & $R_{eff}$ & $\gamma, \theta_{\gamma}$ & Flux ratios &d.o.f. & $\chi^{2}$ & Time delays \\ 
\hline
\hline
(a) HE~0047-1756 & SIS + $\gamma$ & 0.751 & / & / & 0.048, 7.36 & $f_{B}/f_{A}=0.253$ & 0 & 0.0 & $\Delta t_{AB} = 11.9 $\\
& DV + $\gamma$ & 0.756 & 0.22, 113.78 & 0.91 & 0.120, 15.98 & $f_{B}/f_{A}=0.253$ & 0 & 0.0 & $\Delta t_{AB} = 16.5 $\\
\hline
(b) RX~J1131-1231 & SIE + $\gamma$ & 1.834 & 0.20, 117.52 & / & 0.098, 96.37 & $f_{B}/f_{A}=0.615$ & 2 & 200.5 & $\Delta t_{AB} =  -1.0 $ \\
 &  & &  & & &  $f_{C}/f_{A}=0.553$ &  & $\chi^{2}_{im}=66.0$ &  $\Delta t_{AC} = -1.3 $\\
 &  & &  & & &  $f_{D}/f_{A}=0.053$ &  & $\chi^{2}_{l}=120.7$ &  $\Delta t_{AD} = 116.2$\\
& DV + $\gamma$ & 1.791 & 0.32, 114.67 & 1.10 & 0.213, 101.73 & $f_{B}/f_{A}=0.679$ & 3 & 184.6 & $\Delta t_{AB} = -1.7 $\\
 &  & &  &  & & $f_{B}/f_{A}=0.584$ &  & $\chi^{2}_{im}= 56.9$ &  $\Delta t_{AC} = -2.4 $\\
 &  & &  &  & & $f_{B}/f_{A}=0.042$ &  & $\chi^{2}_{l}= 118.4$ &  $\Delta t_{AD} = 198.2 $ \\
\hline
(c) SDSS~J1138+0314 & SIE + $\gamma$ & 0.6640 & 0.05, 118.73 & / & 0.107, 32.12 & $f_{B}/f_{A}=0.505$ & 2 & 2.5 & $\Delta t_{AB} = 3.4 $\\
 &  & &  &  & & $f_{C}/f_{A}=0.714$ &  & $\chi^{2}_{im}=0.1$ &  $\Delta t_{AC} = -1.7 $\\
 &  & &  &  & & $f_{D}/f_{A}=0.945$ &  & $\chi^{2}_{l}=1.9$ &  $\Delta t_{AD} = 0.9 $\\
& DV + $\gamma$ & 0.6629 & 0.15, 121.40 & 0.86 & 0.145, 32.23 & $f_{B}/f_{A}=0.505$ & 3 & 4.7 & $\Delta t_{AB} = 3.8 $ \\
 &  & &  &  & & $f_{C}/f_{A}=0.712$ &  & $\chi^{2}_{im}=0.1$ &  $\Delta t_{AC} = -1.9 $\\
 &  & &  &  & & $f_{D}/f_{A}=0.925$ &  & $\chi^{2}_{l}=4.5$ &  $\Delta t_{AD}= 1.0$ \\
\hline
(d) SDSS~J1155+6346 & SIS + $\gamma$ & 0.59 & / & / & 0.392, 169.66 & $f_{B}/f_{A}=0.710$ & 0 & 0.0 & $\Delta t_{AB} = 20.6 $\\
& DV + $\gamma$ & 0.58 & 0.15, 0.71 & 1.14 & 0.453, 168.98 & $f_{B}/f_{A}=0.710$ & 0 & 0.0 & $\Delta t_{AB} = 25.0$ \\
\hline 
(e) SDSS~J1226-0006 & SIS + $\gamma$ & 0.568 & / & / & 0.100, 8.01 & $f_{B}/f_{A}=0.499$ & 0 & 0.0 & $\Delta t_{AB} = -25.5 $\\
& DV + $\gamma$ & 0.557 & 0.07, 45.18 & 0.69 & 0.145, 4.55 & $f_{B}/f_{A}=0.499$ & 0 & 0.0 & $\Delta t_{AB} = -34.3 $\\
\hline
(f) WFI~J2026-4536 & SIE + $\gamma$ & 0.6520 & 0.22, 167.51 & / & 0.151, 83.46 & $f_{A_{1}}/f_{B}=4.127$ & 2 & 266.7 & $z_{l}$ unknown\\ 
 &  &  &  & & & $f_{A_{2}}/f_{B}=3.439$ &  & $\chi^{2}_{im}=16.3$& \\
 &  &  &  & & & $f_{C}/f_{B}= 1.174$ &  & $\chi^{2}_{l}=66.2$& \\
& DV + $\gamma$ & 0.6517 & 0.25, 167.21 & 0.64 & 0.171, 85.18 & $f_{A1}/f_{B}=3.998$ & 3 & 263.1 & $z_{l}$ unknown\\
 &  &  &  & & & $f_{A_{2}}/f_{B}= 3.307$ &  & $\chi^{2}_{im}=14.4$& \\
 &  &  &  & & & $f_{C}/f_{B}= 1.096$ &  & $\chi^{2}_{l}=62.9$& \\
\hline
(g) HS~2209+1914 & SIS + $\gamma$ & 0.515 & / & / & 0.031, 94.27 & $f_{B}/f_{A}=0.790$ & 0 & 0.0 & $z_{l}$ unknown\\
& DV + $\gamma$ & 0.516 & 0.05, 63.10 & 0.53 & 0.041, 99.60 & $f_{B}/f_{A}=0.790$ & 0 & 0.0 & $z_{l}$ unknown\\
\hline
\end{tabular}
\vspace{0.2cm}
\caption{Results of the parametric modeling.}
\label{lensmodel7}
\end{table*}

\begin{table*}[t!]
\centering 
\begin{tabular}{c||c|cccccc}
\hline
Object & Model & $R_{Ein}$ & $e$ & $\theta_{e}$ & $R_{eff}$ & $\gamma$ & $\theta_{\gamma}$ \\ 
\hline
\hline
(a) HE~0047-1756 & SIS + $\gamma$ & $0.751^{+0.002}_{-0.002}$ & / & / & / & $0.048^{+0.002}_{-0.002}$ & $7.22^{+0.80}_{-0.73}$ \\
& DV + $\gamma$ & $0.755^{+0.003}_{-0.003}$ & $0.22^{+0.01}_{-0.01}$ & $113.33^{+3.92}_{-3.88}$ & $0.91^{+0.01}_{-0.01}$ & $0.119^{+0.005}_{-0.005}$ & $15.66^{+2.09}_{-2.04}$ \\
\hline
(b) RX~J1131-1231 & SIE + $\gamma$ & $1.834^{+0.002}_{-0.002}$ & $0.19^{+0.01}_{-0.01}$ & $117.43^{+0.66}_{-0.63}$ & / & $0.097^{+0.003}_{-0.003}$ & $96.29^{+0.61}_{-0.64}$ \\
& DV + $\gamma$ & $1.790^{+0.002}_{-0.002}$ & $0.31^{+0.01}_{-0.01}$ & $114.60^{+0.52}_{-0.53}$ & $1.10^{+0.02}_{-0.02}$ & $0.212^{+0.002}_{-0.003}$ & $101.66^{+0.22}_{-0.23}$ \\
\hline
(c) SDSS~J1138+0314 & SIE + $\gamma$ & $0.6640^{+0.0005}_{-0.0006}$ & $0.04^{+0.03}_{-0.02}$ & $118.44^{+2.60}_{-3.02}$ & / & $0.105^{+0.006}_{-0.004}$ & $32.20^{+0.37}_{-0.39}$\\
& DV + $\gamma$ & $0.6628^{+ 0.0004}_{-0.0004}$ & $0.15^{+0.01}_{-0.01}$ & $121.16^{+0.94}_{-0.87}$ & $0.86^{+0.02}_{-0.02}$ & $0.145^{+0.001}_{-0.001}$ & $32.24^{+0.26}_{-0.24}$ \\
\hline
(d) SDSS~J1155+6346 & SIS + $\gamma$ & $0.59^{+0.01}_{-0.01}$ & / & / & / &  $0.389^{+0.012}_{-0.015}$& $169.64^{+0.19}_{-0.19}$\\
& DV + $\gamma$ & $0.58^{+0.01}_{-0.01}$ & $0.15^{+0.01}_{-0.01}$ & $0.42^{+2.33}_{-2.45}$ & $1.14^{+0.01}_{-0.01}$ & $0.449^{+0.013}_{-0.014}$ & $168.98^{+0.22}_{-0.21}$ \\
\hline
(e) SDSS~J1226-0006 & SIS + $\gamma$ & $0.568^{+0.003}_{-0.003}$ & / & / & / & $0.100^{+0.005}_{-0.004}$ & $7.94^{+0.38}_{-0.38}$ \\
& DV + $\gamma$ & $0.557^{+0.003}_{-0.003}$ & $0.07^{+0.01}_{-0.01}$ & $44.89^{+4.40}_{-4.42}$ & $0.69^{+0.02}_{-0.02}$ & $0.144^{+0.007}_{-0.007}$ & $4.51^{+0.69}_{-0.70}$ \\
\hline
(f) WFI~J2026-4536 & SIE + $\gamma$ & $0.6518^{+0.0007}_{-0.0006}$ & $0.22^{+0.01}_{-0.01}$ & $167.17^{+0.98}_{-0.88}$ & / & $0.151^{+0.003}_{-0.003}$ & $83.45^{+0.32}_{-0.32}$ \\
& DV + $\gamma$ & $0.6516^{+0.0006}_{-0.0006}$ & $0.24^{+0.01}_{-0.01}$ & $166.98^{+0.74}_{-1.05}$ & $0.64^{+0.01}_{-0.01}$ & $0.171^{+0.002}_{-0.002}$ & $85.21^{+0.22}_{-0.22}$\\
\hline
(g) HS~2209+1914 & SIS + $\gamma$ & $0.515^{+0.002}_{-0.002}$ & / & / & / & $0.031^{+0.002}_{-0.002}$ & $93.89^{+2.00}_{-1.79}$ \\
& DV + $\gamma$ & $0.515^{+0.002}_{-0.002}$ & $0.05^{+0.01}_{-0.02}$ & $63.05^{+2.17}_{-2.10}$ & $0.53^{+0.01}_{-0.01}$ & $0.041^{+0.003}_{-0.003}$ & $96.36^{+4.07}_{-3.20}$ \\
\hline
\end{tabular}
\vspace{0.2cm}
\caption{Median value of the model parameters and 68\% confidence interval.}
\label{MCMC7}
\end{table*}

\section{Discussion}
\label{Discussion7}

\subsection{Doubles}

For the doubly imaged quasars, both SIS+shear and DV+shear models can reproduce the image configuration as well as the flux ratio, even with our constraints on the shape of the galaxy in the case of a de Vaucouleurs profile. Two systems require a large amount of shear (i.e. $\gamma > 0.1$ for both mass models) to reproduce the lens configuration: SDSS~J1226-0006 and SDSS~J1155+6346. For SDSS~J1226-006, the HST/NIC2 images actually reveal a galaxy $\rm G_{2}$ at $\rm RA=1\farcs7153$ and $\rm DEC=3\farcs1710$ from image A (3\farcs4 from the main deflector), about 15$^{\circ}$ off the direction of $\theta_{\gamma}$. This galaxy, which type is unknown, is likely not the only source of shear. Indeed, the luminosity ratio between $\rm G_{2}$ and the lens is $L_{lens}/L_{G_{2}}=4.8$. Assuming we can use the Faber-Jackson relation \citep[L $\propto \sigma^{4}$,][]{Faber1976}, this ratio leads to $\sigma_{lens}/\sigma_{G_{2}}=1.5$, $\sigma_{lens}$ and $\sigma_{G_{2}}$ being respectively the velocity dispersion of the lens and of $\rm G_{2}$. The isothermal model allows us to translate $R_{Ein}$ of the lens to $\sigma_{lens}$. We find $\sigma_{lens}=212$ km/s and thus $\sigma_{G_{2}}=141$ km/s. Using formula A.20 of \citet{Momcheva2006} and supposing $\rm G_{2}$ is at the same redshift as the lens, this induces a shear of $\gamma=0.039$, more than 2 times smaller than the one predicted by the SIS model. Other galaxies in the field are probably responsible for the remaining shear. 

A more dramatic case is SDSS~J1155+6346, for which models predict a shear as large as 0.4 to reproduce the observed configuration. This is one of the largest shears needed to reproduce a lensed quasar system. On some larger field images of this object (obtained with ACS onboard HST, PI: C.S. Kochanek), we do not see any bright galaxy in its vicinity. We thus suspect that a massive galaxy cluster lies outside the ACS field, though nothing is clearly visible on the SDSS data\footnote{\url{http://cas.sdss.org/dr7/en/}}. Deeper images would be necessary to infirm or confirm the existence of this cluster.

In the case of HE~0047-175, a diffuse component lies at $\rm RA=-0\farcs0434$ and $\rm DEC=-2\farcs3393$ from image A ($1\farcs56$ from the lens), in the direction of the shear (see Fig. \ref{dec_NICMOS7}) unregarding the employed model. Although very faint (about 2 mag fainter than the lens), this galaxy is likely the major contribution to the shear in this system. Indeed, a SIS with $\sigma = 88 \ \rm km/s$ would produce the observed amount of shear, if located at the position of this faint companion (assuming $\rm z_{comp} = \rm z_{lens} = 0.407$).

\subsection{Quads and astrometric anomalies}

The quadruply imaged quasars allow to test the ability of simple smooth models to reproduce a relative astrometry with mas accuracy. Only the relative astrometry of SDSS~J1138+0314 is easily reproduced with our models. Conversely, for WFI~J2026-4536 and RX~J1131-1231, we find that a very large $\chi^2$ is associated to our models. In the first case, the main contribution in the $\chi^2$ comes from the difference between the PA of the model and the PA of the light distribution. In the second case, the large $\chi^2$ is mainly due to the impossibility of the model to recover the positions of the lensed images and of the lensing galaxy. In any case, an underestimate of the error bars on the quasar lensed images is unlikely. 

For RX~J1131-1231, a reduced $\chi^2 \sim 1$ can only be obtained if we increase the error bars on the positions of the lensed images by a factor 10. Alternatively, we also get a good fit if we allow more freedom to the position of the lensing galaxy (i.e. error = 0.02). Following this procedure we find that the offset between the light and mass distribution centroid amounts to $88 \rm \ pc$. This value is marginaly consistent with the upper limit of $70 \rm \ pc$ derived for B1938+666 by \citet{Yoo2006}. It is however inconsistent with the maximum offset value of $\sim 20\rm \ pc$ found for the 3 other systems they analysed, suggesting that the offset between light and mass distributions is not the cause of the astrometric perturbation we observe. \citet{Claeskens2006} also found that models (simple or more complex) fitting simultaneously the Einstein ring of RX~J1131-1231 and the quasar lensed images lead to a poor $\chi^2$. \citet{Brewer2008} were also unable to reproduce the lensed quasar relative positions to mas accuracy using the lens model based on the Einstein ring. 

For WFI~J2026-4536, an acceptable $\chi^2$ cannot be obtained in enlarging the error bars on the positions of the lensed images or of the lens galaxy but only in relaxing the constraint on the position angle of this latter. This could be due to the presence of a galaxy located at $\rm RA=-7\farcs398$ and $\rm DEC=-1\farcs940$ \citep{Morgan2004} from image B, that we ignored in the modeling. 

For the three quads, the flux ratios predicted by our best models differ from those measured on the HST/NIC2 frames. This discrepancy may have several origins such as invalid assumption about the lens model, i.e. need for multipole components \citep{Evans2003}, microlensing and/or massive substructures in the lensing galaxies (see \citet{Keeton2003,Keeton2005} for an exhaustive discussion).

The amount of shear needed in our models to reproduce the configuration of the three quads is quite high ($\gamma \gtrsim 0.1$). Our model estimates can be compared to the minimum amount of shear required to reproduce the image position, following the methodology described in \citet{Witt1997}. Using equation (20a) of \citet{Witt1997}, we find a minimum shear $\gamma_{\rm min}$ of resp. 0.035, 0.004 and 0.062 for RX~J1131-1231, SDSS~J1138+0314 and WFI~J2026-4536. The values found for the last two systems are much smaller than the one predicted by our models. This is not surprising as we find $\theta_\gamma$ and $\theta_e$ to be orthogonal, which implies that $\gamma$ is in fact poorly known and strongly degenerate with the internal shear.
For RX~J1131-1231, $\gamma_{\rm min}$ is only 3 times smaller than the value derived for our SIE+$\gamma$ model. The difference between $\gamma_{\rm min}$ and $\gamma_{\rm obs}$ is due to $\theta_e$ being nearly aligned with $\theta_\gamma$ (about 15$^{\circ}$ offset), a situation which also leads to a significant underestimate of $\gamma$ using $\gamma_{\rm min}$. By using eq. (22) from \citet{Witt1997}, we can also derive a range of allowed values for $\theta_e$ based on the image configurations. For WFI~J2026-4536, we find that the observed $\theta_e$ is far from the allowed values, confirming the results of our models which indicate a likely offset betwen the mass and the light matter distribution. For the two other systems, the observed $\theta_e$ falls at the limit of the allowed range, suggestive of a significant ellipticity of the lens, as observed.


The previous results suggest that it is quite common for simple lens models to fail in reproducing mas astrometry of quadruply imaged quasars. To further investigate the question, we have searched the literature for lensed quasars having images with mas astrometric error bars (i.e. up to 0\farcs002 on the lensed image positions) and published simple models. We found eleven systems gathering these conditions\footnote{Although B0712+472 has accurate astrometry, the model published by \citet{Jackson1998} is provided without information about the $\chi^{2}$ thus not allowing to estimate the quality of the fit.}: B0128+437 \citep{Biggs2004}, MG0414+0534 \citep{Ros2000}, HE0435-1223  \citep{Morgan2005,Kochanek2006}, SDSS0924+0219 \citep{Keeton2006, Cosmograil2}, H1413+117 \citep{chantry2007, MacLeod2009}, B1422+231 \citep{Bradac2002},  B1608+656 \citep{Koopmans2003}, B1933+503\footnote{B1933+503 is actually a ten-images lens. Only the position of four of them is known with a precision of 2 mas or less.} \citep{Cohn2001}, MG2016+112 \citep{Chen2007,More2009}, WFI2033-4723 \citep{Cosmograil7}, B2045+265 \citep{McKean2007}. Two of these systems (H1413+117, B1933+503) are easily reproduced by simple models because of the large uncertainty affecting the position of the lensing galaxy ($\sigma_{gal}>0\farcs01$). Out of the six systems for which VLA, VLBA or VLBI data are available\footnote{B1933+503 also has VLA data but has already been ruled out, its configuration being easily reproduced with simple models because of the large uncertainty on the lens position.} (B0128+437, MG0414+0534, B1422+231, B1608-656, MG2016+112, B2045+265), only B1422+231 shows convincing evidence that smooth models allow to reproduce the relative astrometry, although substructures are needed to reproduce the flux ratios \citep{Bradac2002}. For MG0414+0534, B1608+656, MG2016+112 and B2045+265, complex models including a bright susbtructure (MG0414+0534) or a companion galaxy (B1608+656, MG2016+112, B2045+265) are needed to get to acceptable fits. The case of B0128+437 is a bit peculiar as the lensing galaxy was not detected at the time of the modeling paper\footnote{The latter has been unveiled recently on AO images by \citet{Lagattuta2010}.} and so, only the lensed images were used for the models, the latter being constrained as more than four lensed images are detected. Based only on the position of the lensed images, an astrometric anomaly is detected as long as an astrometric accuracy smaller than 1 mas (but still compatible with the data error bars) is considered. Out of the last three systems (HE0435-1223, SDSS0924+0219, WFI2033-4723), constrained by relative astrometric positions derived from HST images, a good fit is obtained only for HE0435-1223. A SIE+$\gamma$ model leads to $\chi^{2}_{r}\sim33$ for SDSS0924-0219 \citep{Keeton2006} and to $\chi^{2}_{r}\sim15$ for WFI2033-4723. In both cases, relaxing the constraint on the lens galaxy centroid allows to get a perfect fit of the astrometry. For WFI2033-4723 more complex models including a nearby group allow to reproduce the astrometry but they also need the light and mass distribution to be misaligned, which is not totally satisfactory \citep{Cosmograil7}. \citet{KochanekDalal2004} compile seven quads (some of them also compiled here) for which they fitted SIE+$\gamma$ models. Unfortunately, there is only sparse information on the astrometric error used and we cannot infer any trend from this study. 

To conclude, out of the nine usable systems amongst the eleven quads, at least four show astrometric perturbations with respect to predictions of simple lens models (B0128+437, MG0414+0534, SDSS0924+0219, WFI2033-4723). For three of the remaining systems (B1608+656, MG2016+112, B2045+265), conclusions are difficult to draw because the need to include a companion galaxy comes naturally from deep near-IR imaging. The last two systems (HE0435-1223, B1422+231) are well reproduced by simple smooth models. 

Although the considered sample of quads gathers heterogeneous data sets and analyses, it indicates that relative astrometry of quads often deviates from simple models expectation when trying to reproduce it to the mas precision. The considered sample suggests that the situation is less critical for \textquotedblleft central quads\textquotedblright \ (i.e. with the source lying close to the center of the central astroid caustic) than for fold systems (i.e. source lying close to a fold caustic). This might be a normal geometrical effect (image positions vary more slowly when moving the source in the central region of the astroid caustic) but it might also be due to substructures leading to severe deformations of the caustics \citep{Bradac2002}. It remains to be seen how significant this effect is with respect to the relative astrometric uncertainty on the image positions or the amount of shear.

Substructures are not the only explanation for the frequent inability of simple lens models to fit the configuration of quads. Other possible explanations are astrometric perturbations due to the lens environment, asymmetries in the mass distribution, disky/boxy projected mass profiles, offsets between the galaxy light centroid and mass centroid. The last two solutions seem however to be ruled out by \citet{KochanekDalal2004} and \citet{Yoo2006}. The evidence that bright substructures/nearby satellite galaxies explain astrometric perturbations of some systems suggest that substructures may be one of the major contributors to the astrometric perturbations of quads. 

All this motivates a systematic study of the ability of simple models to reproduce the configuration of quads, with good control on the error estimates and uniform modeling. Such a work is beyond the scope of this paper and is delayed to a forthcoming paper, when the iterative deconvolution method will have been applied to a larger number of quadruply imaged quasars.

\section{Conclusions}
\label{conc7}
In applying ISMCS, i.e. the MCS deconvolution algorithm combined with an iterative strategy, to HST/NIC2 images of seven lensed quasars, we have obtained accurate relative positional constraints on the lensed images, lensing galaxy as well as shape parameters for this galaxy. We reach an accuracy of around 1-2 mas on the lensed image positions. We also detect for the first time a partial Einstein ring in two cases, the double HE~0047-1756 and the quadruple SDSS~J1138+0314, and we highlight as well the already known ring in RX~J1131-1231. Deeper images are needed to perform clear source reconstruction. In the case of HS~2209+1914, the deconvolved frame reveals a structure around the bulge of the lens galaxy which cannot be clearly identified. This structure could be either an Einstein ring or some not well-resolved spiral arms of a big late-type galaxy. This question probably deserves further study: a spectrum of the surrounding structures could give extra information about their true nature. 

We also obtain simple mass models for every system. In the case of doubles, both the isothermal and de Vaucouleurs profiles can reproduce the observed configuration. For SDSS~J1155+6346, a good fit can only be reached with an extremely and anomalously high external shear, 0.392 in the case of a SIS and 0.453 in the case of a DV, indicating the presence of a galaxy group or cluster probably located outside the field of view of the ACS. For SDSS~J1226-006, the large shear ($\gamma$=0.1) is probably partially due to a nearby galaxy located 3\farcs4 from the main deflector. In the case of quads, a good $\chi^{2}$ can only be obtained for one object: SDSS~J1138+0314. The two other quads of our sample, RX~J1131-1231 and WFI~J2026-4536, need more complicated models to account for their observed configuration. For RX~J1131-1231, the offset between the light and mass distribution cannot account for the astrometric perturbation we observe. 

The study of the literature allows us to conclude that most of the quads cannot be modeled with simple profiles when the astrometic accuracy reaches around 1 mas: some need the presence of companion galaxies, some others need substructures. This finding motivates the acquisition of mas astrometry for all the quads together with simple modeling.

In the framework of the COSMOGRAIL collaboration, the next step for these seven systems is the acquisition of well-sampled light curves to extract time delays. Then, if the redshift of the lens is known, our astrometric constraints will help in reducing the systematic errors on the Hubble constant. 

\begin{acknowledgements}
COSMOGRAIL is financially supported by the Swiss National Science Foundation (SNSF). This work is also supported by the Belgian Federal Science Policy (BELSPO) in the framework of the PRODEX Experiment Arrangement C-90312. VC thanks the Belgian National Fund for Scientific Research (FNRS) as well as ESO Vitacura (Chile) for the use of their facilities. A fellowship from the Alexander von Humboldt Foundation to DS is gratefully acknowledged. We also thank Ross Fadely and Chuck Keeton for advices on the use of the MCMC implementation of LENSMODEL. And finally we thank the anonymous referee for many precious comments which allowed to improve this paper.
\end{acknowledgements}

\bibliographystyle{aa}
\bibliography{/datas5/Programmes/JabRef/Articles}

\end{document}